\documentclass[a4paper,11pt]{article}
\pdfoutput=1 % for arXiv: compile with pdfLaTeX

\usepackage{jheppub} % JHEP style; loads hyperref with JHEP's link settings

\usepackage{amsmath,amssymb,amsthm,mathrsfs,bbm}
\usepackage{latexsym,amscd,amsbsy,amsfonts,dsfont}
\usepackage{graphicx}
\usepackage{physics}
\usepackage{comment}
\usepackage[dvipsnames]{xcolor}
\usepackage{tikz}

\renewcommand{\dd}{\mathrm{d}}
\renewcommand{\L}{\mathcal{L}}

\numberwithin{equation}{section}

\title{Challenging the hoop conjecture with spinning pancakes and baguettes}

\author[a,b]{Robin Croft,}
\author[a,b]{Gerardo Garc\'ia-Moreno,}
\author[a,b]{and Paolo Pani}

\affiliation[a]{Dipartimento di Fisica, Sapienza Universit\`a di Roma, Piazzale Aldo Moro 5, 00185, Roma, Italy}
\affiliation[b]{INFN, Sezione di Roma, Piazzale Aldo Moro 2, 00185, Roma, Italy}

\emailAdd{robin.croft@uniroma1.it}
\emailAdd{gerardo.garciamoreno@uniroma1.it}
\emailAdd{paolo.pani@uniroma1.it}

\abstract{Understanding when gravitational collapse produces a black hole remains a central challenge in general relativity. Black-hole uniqueness results characterize the stationary endpoint under suitable assumptions, while singularity theorems establish geodesic incompleteness in the presence of trapped surfaces and appropriate energy and global conditions. Neither, however, provides a general criterion for horizon formation. The hoop conjecture offers such a criterion by relating the enclosed mass to the compactness of a matter distribution in every direction. We construct axisymmetric families of initial data describing rotating dust and test a formulation of the conjecture adapted to this setting. We explore highly prolate and oblate configurations with both constant and differential rescaled angular-velocity profiles, including configurations near the threshold for apparent-horizon formation. We find that the conjecture holds for rotating configurations, with a coefficient within $14\%$ of unity as in the nonrotating case, provided the quasilocal mass includes the rotational energy. If the irreducible mass of the apparent horizon is used instead, the deviations grow with spin, as they do for Kerr black holes.
For horizonless configurations, the conjecture likewise holds when the mass is restricted to the energy enclosed by the matter surface, whereas using the ADM mass, which also includes rotational field energy outside the matter, leads to apparent violations of up to $\simeq 25\%$. We also discuss the possible evolution of compact configurations without apparent horizons, their relation to weak cosmic censorship, and their implications for horizonless compact objects.}


\begin{document}
%--------------------------------------------------------------------

\maketitle
\flushbottom

\section{Introduction}
Understanding how gravitational collapse proceeds within General Relativity (GR) remains a central open problem. A central question is the nature of the final state reached by a collapsing system. The no-hair results suggest that, provided a black hole forms and the system settles down, its exterior geometry should ultimately approach that of a member of the Kerr family~\cite{Chrusciel2012}. At the same time, the formation of a trapped surface has profound implications for the global structure of spacetime: under suitable assumptions, including global hyperbolicity and appropriate energy conditions, the singularity theorems ensure geodesic incompleteness~\cite{Penrose1964,Penrose1969,Senovilla2014}. These results, however, do not by themselves determine the dynamical conditions leading to horizon formation. In particular, establishing when a given matter configuration develops an apparent horizon remains a challenging and largely open aspect of gravitational collapse.

\begin{figure}[htbp]
    \centering
    \includegraphics[width=0.4\textwidth]{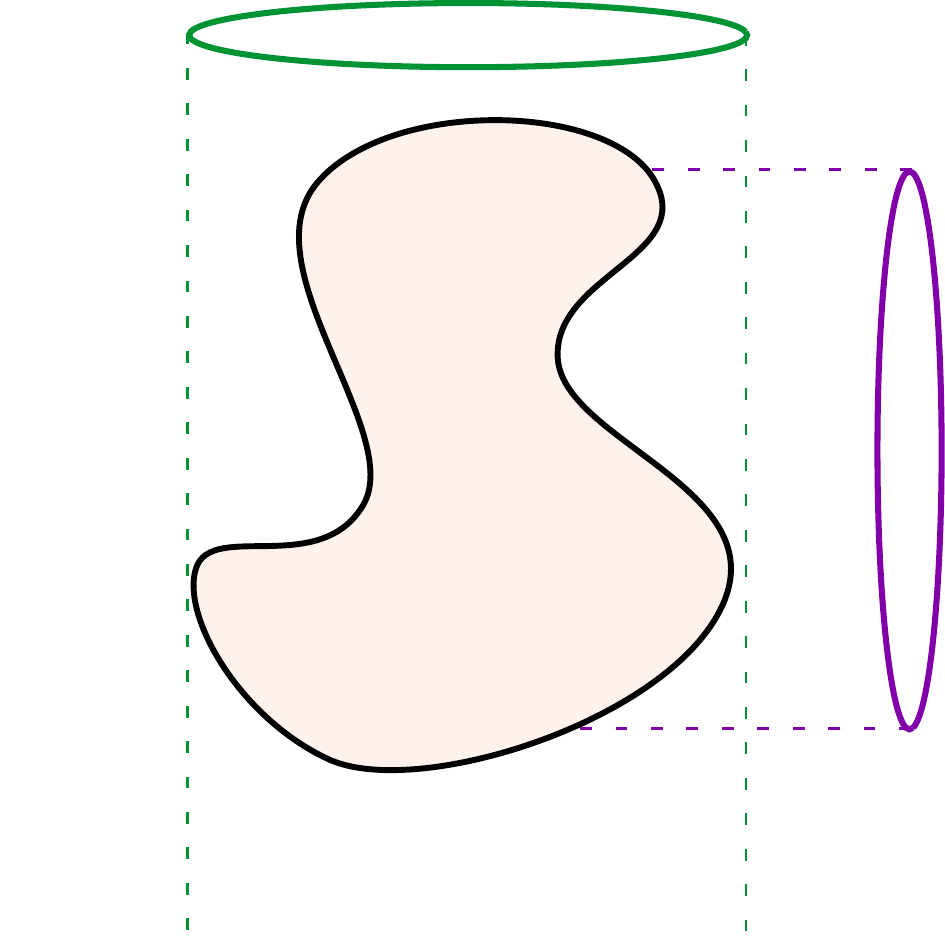} \hspace{1cm}
    \includegraphics[width=0.4\textwidth]{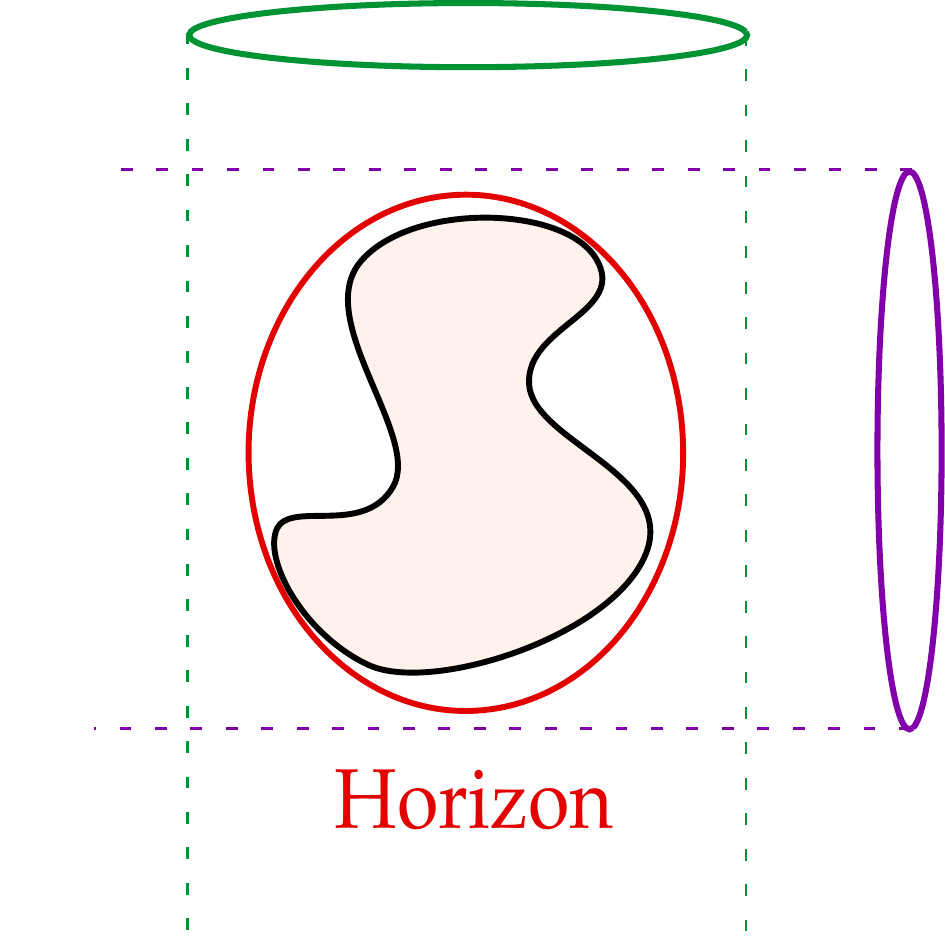}
    \caption{Illustration of Thorne's hoop conjecture for a nonsymmetric matter distribution of mass $M$. Hoops have circumference $4\pi M$.
    Left: a hoop can be placed around the matter in the vertical plane (green), but not in the horizontal plane (purple). The conjecture therefore predicts that no black hole forms.
    Right: hoops of circumference $4\pi M$ fit around the matter in \emph{every} direction, so a horizon (red) is expected.}
    \label{Fig:PictorialHoop}
\end{figure}

In that sense, one of the most important proposals related to this is the hoop conjecture proposed by Thorne in 1972~\cite{Thorne1972}. Motivated by the study of a toy model for an anisotropic collapse, Thorne proposed it as a tentative criterion to determine under which conditions a configuration would form a black hole. He suggested that a \emph{black hole forms if and only if a hoop of circumference $4 \pi M$, where $M$ represents the mass that is enclosed in the region, can be passed around the matter in every direction, i.e., $\mathcal{C}_{\text{min}}  \lesssim 4 \pi M $ where $\mathcal{C}_{\text{min}} $ represents the minimum size of a hoop that passes through every direction.} See Fig.~\ref{Fig:PictorialHoop} for a pictorial representation.

The conjecture contains two implications. The \emph{if} part states that if a hoop of circumference $\mathcal{C} \lesssim 4\pi M$ can be placed around the matter in every direction, then a horizon forms. Equivalently, if no horizon is present, there must be at least one direction in which no such hoop fits around the matter. We can test this on configurations without a horizon by checking that at least one minimal enclosing hoop exceeds $4\pi M$. The \emph{only if} part is the converse: a horizon forms only if such hoops fit in every direction. We can test this on configurations with a horizon by checking that the horizon circumferences satisfy $\mathcal{C} \lesssim 4\pi M$ in every direction.

Stated this way, however, the conjecture is ambiguous, and making it precise has been an active area of research; see, e.g.,~\cite{Flanagan1991,Senovilla2007,Wald1997} and references therein, as well as Refs.~\cite{Yoo2016,East2019,Ames2023,Patino:2026jlq} for recent numerical studies of highly compact configurations near the threshold of collapse. The main difficulties are threefold. First, energy is not localisable in GR, so it is not obvious how to define the mass contained in a region. Second, the notion of horizon is subtle: event horizons are teleological, while trapped surfaces depend on the slicing. Third, the proportionality constant in the inequality is unspecified, and it may even differ between the \emph{if} and \emph{only if} parts of the conjecture~\cite{Flanagan1991}. A more recent proposal~\cite{Gibbons2009,Cvetic2011} avoids the first two issues by bounding a geometric invariant of the apparent horizon, its Birkhoff invariant, in terms of the ADM mass, and verifies this bound for several classes of black holes, including Kerr.

Here we focus on a setup in which the conjecture can be formulated sharply. We consider axisymmetric spheroids of rotating dust, either prolate (``baguettes'') or oblate (``pancakes''), and solve the GR constraint equations on a Cauchy slice. We build on the pioneering work of Nakamura, Shapiro, and Teukolsky~\cite{Nakamura1988}, extending their framework to rotating dust. The role of rotation in the hoop conjecture has so far been studied mainly for stationary black holes~\cite{Senovilla2007,Gibbons2009,Cvetic2011}. To our knowledge, it has not been explored for dynamical, matter-filled initial data near the threshold of horizon formation. Our aim is to provide a simple setting in which this question can be addressed quantitatively.

In particular we find a good agreement with most of previous results. First of all, we see that in the cases in which we find a configuration with horizons, the \emph{only if} part of the hoop conjecture is satisfied within $14 \%$ and hence it captures the necessary
conditions for the formation of the horizon, up to a numerical coefficient. However, we corroborate that it does not seem possible to provide a universal coefficient for the \emph{if} part of the conjecture. 

The addition of rotation introduces the novelty of needing to account for the rotational energy of the configuration. In the axisymmetric case, we have two natural candidates for the mass of a horizon, one that includes a contribution from rotation, which we dub the total mass, and another one that does not, which is given by the area of the horizon. For our most rapidly spinning configuration, the horizon satisfies the conjecture when the total mass is used, but violates it by $\simeq6\%$ when the irreducible mass is used. More generally, with the total mass, rotation does not increase the deviations beyond those already induced by eccentricity in the nonrotating case. Although this $6\%$ violation is smaller than the eccentricity-induced deviations, the shift between the two masses depends only on the horizon spin and grows with it.
For Kerr black holes it reaches a factor $\sqrt{2}$ at extremality, see Eq.~\eqref{Eq:KerrExtremalHoop} and Ref.~\cite{Senovilla2007}. This indicates that the mass entering the hoop conjecture should include the rotational energy. Larger spins would make the difference more pronounced, but cannot be reached within our setup without the dust becoming superluminal. Exploring matter models that allow faster rotation is therefore a natural extension of this work.

For the \emph{if} part, the outcome depends on how the rotational energy is assigned. In the rotating case, the ADM mass is an integral over the whole slice and includes energy outside the matter. If, consistently with the horizon case, we use the energy enclosed by the surface on which the hoop is measured, namely the contained mass $M_{\text{cont}}$ obtained by restricting the integral to the matter region, the conjecture is satisfied in all our horizonless configurations. If the full ADM mass is used instead, the minimal hoops can be up to $\simeq 25\%$ smaller than $4\pi M_{\text{ADM}}$ in the absence of a horizon. This ambiguity is specific to rotation: without rotation, $M_{\text{ADM}} = M_{\text{cont}}$ in our setup.

Here is an outline of the remainder of the article. In Sec.~\ref{Sec:InitialData} we construct the initial data that we are going to use to test the hoop conjecture; in Subsec.~\ref{Subsec:3+1} we present the 3+1 decomposition for rotating dust that we are considering, in Subsec.~\ref{Subsec:Asymptotics} we examine the asymptotic behaviour of the solutions to find suitable boundary conditions and to obtain integral expressions for the ADM quantities of the spacetime, and in Subsec.~\ref{Subsec:NumericalImplementation} we discuss our numerical implementation to solve the constraints. In Sec.~\ref{Sec:Horizons} we analyze under which conditions the initial data that we have constructed exhibit a horizon, i.e. a marginally outer trapped surface~(MOTS), in Subsec.~\ref{Subsec:ApparentHorizons} we derive the equations that MOTS should satisfy in our setting, in Subsec.~\ref{Subsec:MassAngularMomentum} we give the expressions for the mass and angular momentum of the horizon and we present our numerical scheme to compute them in Subsec.~\ref{Subsec:NumericalHorizon}. In Sec.~\ref{Sec:Hoops} we revisit the hoop conjecture, we discuss the motivation for formulating it in Subsec.~\ref{Subsec:ReviewHoop}, we explain how the problem of making it mathematically precise can be overcome in our setting in Subsec.~\ref{Subsec:AxialSym} and in Subsec.~\ref{Subsec:OverSpinning} we discuss the potential end-state of the collapse in the case in which no horizon is formed. In Sec.~\ref{Sec:Numeric} we present our numerical results, first we consider the case in which the fluid is not rotating in Subsec.~\ref{Subsec:non-rotatingConfigs}, which is quite similar to previous analysis in the literature, we consider rigidly rotating dust in Subsec.~\ref{Subsec:RigidRotation} and differentially rotating dust in Subsec.~\ref{Subsec:DifferentialRotation}. In Sec.~\ref{Sec:Conclusions}, we summarise the main results of the article and we present our conclusions together with possible extensions of this work. 

Some computational details are collected in the appendices. App.~\ref{App:ExtrinsicCurvature} details the York--Lichnerowicz decomposition, App.~\ref{App:HorizonAppendix} derives the boundary conditions used in the shooting method, App.~\ref{App:NearHorizon} shows how the relaxation-flow algorithm converges to the Schwarzschild horizon for nearby initial guesses, and App.~\ref{App:Convergence} presents the numerical convergence 
of the multigrid solver.

%-------------------------------------------------------
 \paragraph*{\textbf{Notation and conventions.}}
%-------------------------------------------------------

We use the signature $(-,+,+,+)$  for the spacetime metric and we work in geometrised units in which $c=G_N=1$. For the curvature tensors we use the conventions in the book of Wald~\cite{Wald1984}, i.e., $[\nabla_\mu, \nabla_\nu] V^\rho =: - R_{\mu \nu \sigma}{}^\rho V^\sigma$, $R_{\mu \nu}:=R_{\mu \nu \rho}{}^\rho$. We use Greek indices as indices in the 4-dimensional manifold $(\mu, \nu...)$ and latin indices $(i,j...)$ for indices within the 3-dimensional slices of the 3+1 decomposition. We also introduce $H_{(ab)}:= \frac{1}{2!}(H_{ab}+H_{ba})$ and $H_{[ab]}:= \frac{1}{2!}(H_{ab}-H_{ba})$, and similarly for an object with $n$ indices instead of 2.

\section{Constructing the initial data}
\label{Sec:InitialData}
We solve the GR initial constraints for a self-gravitating axisymmetric rotating cloud of dust. These families of solutions constitute a priori suitable data on a given Cauchy slice of the spacetime and we can use them to test the hoop conjecture. In Subsec.~\ref{Subsec:3+1} we present the $3+1$ decomposition for this setup and write down the PDEs that we need to solve, in Subsec.~\ref{Subsec:Asymptotics} we discuss the asymptotic behaviour of the functions and extract the ADM mass and angular momentum, and in Subsec.~\ref{Subsec:NumericalImplementation} we discuss our numerical implementation to solve these PDEs.

\subsection{3+1 decomposition}
\label{Subsec:3+1}
We introduce the $3+1$ decomposition of the spacetime:
\begin{align}
    \dd s^2 = g_{\mu \nu} \dd x^{\mu} \dd x^{\nu} = - \alpha^2 \dd t^2 +  \gamma_{ij} \left( \dd x^i + \beta^i \dd t \right)  \left( \dd x^j + \beta^j \dd t \right), 
\end{align}
in terms of which the future directed unit timelike vector normal to the constant $t$ surfaces is given by:
\begin{align}
    n^{\mu} = - \alpha g^{\mu \nu} \partial_{\nu} t, 
    \label{Eq:n_vector}
\end{align}
or in components 
\begin{align}
    n^\mu &=  \alpha^{-1}[(\partial_t)^\mu - \beta^\mu],\\
    n_\mu &=- \alpha \left( \dd t \right)_{\mu} . 
\end{align}
In terms of $n^{\mu}$, the extrinsic curvature of the constant $t$-surfaces can be expressed as:
\begin{align}
    K_{ij} = - \frac{1}{2} \L_{n} \gamma_{ij} .
\end{align}
The projector onto the constant $t$-slices is given by
\begin{align}
    P^{\mu}_{\ \nu} = \delta^{\mu}_{\ \nu} + n^{\mu} n_{\nu}.
\end{align}
The Hamiltonian and momentum constraints on a constant time slice with induced metric $\gamma_{ij}$ and extrinsic curvature $K_{ij}$ are:
\begin{align}
    & \ ^3{R} (\boldsymbol{\gamma}) + K^2 - K_{ij}K^{ij} = 16 \pi \rho, \label{Eq:HamConstraint0} \\
    & D_{j} \left( K^{ij} - \gamma^{ij} K \right) = 8 \pi j^{i}, \label{Eq:MomConstraints}
\end{align}
where $D_i$ represents the torsionless covariant derivative compatible with $\gamma_{ij}$, $K$ represents the trace of the extrinsic curvature $K:=\gamma^{ij}K_{ij}$, $^3{R} (\boldsymbol{\gamma})$ represents the Ricci scalar of the three-dimensional metric $\boldsymbol{\gamma}$ and
\begin{align}
    \rho & : = T_{\mu \nu} n^{\mu} n^{\nu}, \\
    j^{i} & := - T_{\mu \nu} n^{\mu} P^{\nu i},
\end{align}
are the energy density and momentum density of matter perceived by local observers following integral curves of $n^{\mu}$. 

We restrict the analysis to axisymmetric and conformally flat spatial metrics, i.e., in suitable cylindrical-like coordinates $(r,\phi,z)$ we have:
\begin{align}
    \dd s_3^2 = \psi^4( r, z) \left( \dd r^2 + r^2 \dd \phi^2 + \dd z^2 \right).
\end{align}
The physical metric $\gamma_{ij} $ and the flat one $\tilde{\gamma}_{ij}$ are related through $ \gamma_{ij} = \psi^4( r, z) \tilde{\gamma}_{ij}$. We will use the tilde to represent the objects defined with respect to the flat metric, e.g., $\tilde{D}_i$ will represent the torsionless covariant derivative compatible with $\tilde{\gamma}_{ij}$.

We assume that the slicing is maximal, meaning that $K = 0$. The resulting traceless extrinsic curvature can be decomposed in terms of the York-Lichnerowicz decomposition as:
\begin{align}
    K^{ij} = \psi^{-10} \tilde{A}^{ij} = \psi^{-10} \tilde{A}_{\text{TT}}^{ij} + \psi^{-10}\left(\boldsymbol{\tilde{L}} W \right)^{ij},
\end{align}
where $\tilde{A}_{\text{TT}}^{ij} $ is the transverse traceless part of the tensor $\tilde{A}_{ij}$ whereas the vector part $W^i$ appears in the equation through the Lichnerowicz operator:
\begin{align}
    \left(\boldsymbol{\tilde{L}} W \right)^{ij} = 2 \tilde{D}^{(i} W^{j)} - \frac{2}{3} \tilde{\gamma}^{ij} \tilde{D}_k W^k.  
\end{align}
This decomposition requires the choice of a background metric with respect to which we can define the Lichnerowicz operator. We choose the flat metric as it is more convenient for our purposes, hence we have denoted the operator with a tilde $\boldsymbol{\tilde{L}}$. We provide the details of the computation of $\left(\boldsymbol{\tilde{L}} W \right)^{ij} $ in App.~\ref{App:ExtrinsicCurvature}. 

We emphasise that $\tilde{A}_{\text{TT}}^{ij}$ is transverse, i.e., divergenceless with respect to the conformal connection (whose derivative operator is $\tilde{D}_i$), consequently it is not fixed by the constraint equations, since it automatically drops from the momentum constraint~\eqref{Eq:MomConstraints}, and the Hamiltonian constraint~\eqref{Eq:HamConstraint0} does not contain any derivatives of it. It is often interpreted as encoding the presence of gravitational radiation in the initial data. As such, we take $\tilde{A}_{\text{TT}}^{ij} = 0$ to focus on the role of matter and its rotation in gravitational collapse. Furthermore, the hoop conjecture invokes the energy and mass content within a given region. Had we added a nontrivial gravitational wave content that is not determined by the matter distribution itself, we would have had to account for its energy which is beyond the scope of this work.

These assumptions together simplify the equations to:
\begin{align}
    & \tilde{\Delta} \psi + \frac{1}{8} \psi^{-7} \left( \boldsymbol{\tilde{L}} W \right)^{ij} \left( \boldsymbol{\tilde{L}} W \right)_{ij} = - 2 \pi \psi^5 \rho , \label{Eq:HamConstraint} \\
    & \tilde{D}_j \left( \boldsymbol{\tilde{L}} W \right)^{ij} = 8 \pi \psi^{10} j^{i}. \label{Eq:MomConstraint}
\end{align}
where here $\tilde{\Delta}$ represents the Laplacian operator with respect to the flat metric $\tilde{\gamma}_{ij}$. 

Now we need to specify the matter content. We consider a pure dust energy-momentum tensor:
\begin{align}
    T_{\mu \nu} = \rho_{D} u_{\mu} u_{\nu},
\end{align}
with $\rho_{\text{D}}$ representing the density of the dust in the comoving frame and $u^{\mu}$ the four-velocity of the fluid. Furthermore, we assume that the fluid is only moving in the $\phi$ direction, i.e., it is rotating with respect to the $z$-axis which is $r = 0$. This means that the energy-momentum tensor of the fluid only has momentum in that direction too. To be more specific, the four velocity of the fluid is given by:
\begin{align}
    u^{\mu} = \frac{1}{\sqrt{1-\left( v^{\phi}\right)^2 r^2 \psi^4}} \left( n^{\mu} + v^{\mu}\right)
\end{align}
where the prefactor is ensuring that $u^{\mu}$ is normalized, i.e., $u_{\mu} u^{\mu} = -1$, whereas $v^{\mu} \neq 0$ indicates that it is not aligned with the time direction $n^{\mu}$, and we have assumed that it only has a component in the $\phi$ direction
\begin{align}
    v^{\mu} = v^{\phi} \left( \partial_{\phi} \right)^{\mu}.
\end{align}
We note that the vector component $v^{\phi}$ is an angular velocity.

In terms of this energy-momentum tensor, the density and momenta in the right hand side of Eqs.~\eqref{Eq:HamConstraint}-\eqref{Eq:MomConstraint} are given by:
\begin{align}
    \rho & = T_{\mu \nu} n^{\mu} n^{\nu} =  \frac{\rho_{\text{D}}}{1-\left(v^{\phi}\right)^2 r^2 \psi^4} ,\\
    j^i &  = - T_{\mu \nu}  P^{\mu i}  n^{\nu} = \frac{\rho_{\text{D}}}{1-\left(v^{\phi}\right)^2 r^2 \psi^4} v^{\phi} \delta^{i}_{\ \phi} . 
\end{align}
Notice that the only nonvanishing component of $j^{i}$ is $j^{\phi}$. This also simplifies the vector $W^i$ from the Lichnerowicz decompositions and the equations that it satisfies. First of all, the only nonvanishing component is $W^{\phi}$ or, equivalently, $W_{\phi}$, as the metric is diagonal. Furthermore, the only of the three momentum constraints that is not automatically satisfied leads to the following PDE:
\begin{align}
    \left(\partial_r^2 +\frac{3}{r}\partial_r + \partial_z^2 \right) W^\phi = 8 \pi \psi^{10} \frac{\rho_{\text{D}}}{1-\left(v^{\phi}\right)^2 r^2 \psi^4} v^{\phi}. 
\end{align}
We now introduce the convenient change of variables $\mathcal{W}(r,z) := r  W^{\phi} $, in terms of which we can rearrange the second equation as:
\begin{align}
    \left(\partial_r^2 +\frac{1}{r}\partial_r + \partial_z^2 - \frac{1}{r^2} \right) \mathcal{W}(r,z) = 8 \pi \psi^{10} r \frac{\rho_{\text{D}}}{1-\left(v^{\phi}\right)^2 r^2 \psi^4} v^{\phi}. 
\end{align}
Putting everything together, this is the PDE system that we need to solve to find initial data:
\begin{align}
    & \left( \partial_r^2 + \frac{1}{r} \partial_r + \partial_z^2 \right) \psi + \frac{1}{4} \psi^{-7}   \left[   \left( \partial_r  \mathcal{W} \right)^2 + \frac{ \mathcal{W}^2}{r^2} - 2 \frac{ \mathcal{W} \partial_r  \mathcal{W}}{r} +  (\partial_z  \mathcal{W})^2 \right]  \nonumber \\
    & = - 2 \pi \psi^5 \frac{\rho_{\text{D}}}{1-\left(v^{\phi}\right)^2\psi^4 r^2 }\,, \label{Eq:PsiPhysequation} \\
    & \left(\partial_r^2 +\frac{1}{r}\partial_r + \partial_z^2 - \frac{1}{r^2} \right) \mathcal{W}(r,z) = 8 \pi \psi^{10} r \frac{\rho_{\text{D}}}{1-\left(v^{\phi}\right)^2 r^2 \psi^4} v^{\phi}. \label{Eq:WPhysequation}
\end{align}
These equations can be solved for $\psi$ and $W^{\phi}$, once the functions $\rho_{D}$ and $v^\phi$, encoding the initial distribution of the fluid are specified. Both equations are nonlinear and they are also coupled due to the $\psi$ factors appearing on the right hand sides. 

A priori, there is no guarantee that these PDEs admit a solution for a given arbitrary profile $\rho_{\text{D}}$ and $v^{\phi}$, even if the density is compactly supported. In fact, even for nonrotating dust $v^{\phi} = 0$, which leads to $\mathcal{W}=0$, it is known that the remaining simpler PDE:
\begin{align}
    \tilde{\Delta} \psi = - 2 \pi \psi^5 \rho_{\text{D}},
\end{align}
does not always admit a solution. In particular, if the density $\rho_{\text{D}}$ becomes sufficiently large in a region, the solution ceases to exist. An explicit example is worked out in Ref.~\cite{Baumgarte2006}, where this equation is solved analytically for a spherically symmetric density profile. In particular, it is taken to be the characteristic function inside of a sphere of radius one\footnote{We define the characteristic function of a set as the function that is $1$ when evaluated on points of the set and $0$ for points that do not belong to it.}. There is a critical value, $\rho_0$, beyond which the equation does not admit solutions. Furthermore, for the values for which there exists a solution, it turns out that it is not unique. In the particular example discussed in that reference, there are two solutions for each value of $\rho_0$ below the critical one. The authors argue that one of the two branches can be excluded on physical grounds, as it is a solution for which the conformal factor $\psi$ grows unboundedly as the $\rho_0 \to 0$ limit is approached. 

We interpret the absence of solutions for sufficiently large densities as having a physical origin. Our data describe a snapshot of the spacetime on a maximal slice in which the dust carries no radial momentum, and they can be evolved both forward and backward in time. Nonexistence then suggests that GR does not allow an instant at which an arbitrarily large amount of matter is concentrated within a given region while being momentarily neither expanding nor collapsing.

This means that for a given specified $\rho_{\text{D}}$ and $v^{\phi}$ profiles, we do not know a priori if there is a solution to the initial data constraints. To keep the setup as simple as possible, we have decided to stick to the framework in which there is no radial momentum but follow the idea presented in Ref.~\cite{Nakamura1988}. We can specify $\rho_{\text{N}}$ and $v^{\phi}_{\text{N}}$, instead of $\rho_{\text{D}}$ and $v^{\phi}$, which are the following combination of functions 
\begin{align}
    & \rho_{\text{N}} :=  \psi^5 \frac{\rho_{\text{D}}}{1-\left(v^{\phi}\right)^2\psi^4 r^2 }, \label{Eq:rhoNdef}\\
    &  v_{\text{N}}^{\phi} := \psi^5 v^{\phi}, \label{Eq:vNdef}
\end{align}
which appear on the right hand side of Eqs.~\eqref{Eq:PsiPhysequation} and~\eqref{Eq:WPhysequation}. This simplification transforms Eq.~\eqref{Eq:WPhysequation} into a linear equation that can be straightforwardly solved and always admits a solution for compactly supported density profiles $\rho_{\text{N}}$. After integrating it, the result can be plugged into Eq.~\eqref{Eq:PsiPhysequation} which, although still not linear, does not suffer from the existence and uniqueness problem. In terms of these sources, the PDEs that we solve are:
\begin{align}
    & \left( \partial_r^2 + \frac{1}{r} \partial_r + \partial_z^2 \right) \psi + \frac{1}{4} \psi^{-7} \left[   \left( \partial_r  \mathcal{W} \right)^2 + \frac{ \mathcal{W}^2}{r^2} - 2 \frac{ \mathcal{W} \partial_r  \mathcal{W}}{r} +  (\partial_z  \mathcal{W})^2 \right]  = - 2 \pi \rho_{\text{N}},  \label{Eq:PsiRescequation}\\
    &  \left(\partial_r^2 +\frac{1}{r}\partial_r + \partial_z^2 - \frac{1}{r^2} \right) \mathcal{W}(r,z) = 8 \pi r \rho_{\text{N}} v_{\text{N}}^{\phi}, \label{Eq:WPRescequation}
\end{align}
where we recall that we are specifying the functions $\rho_{\text{N}},v_{\text{N}}^{\phi}$ together with the boundary conditions that the function $\mathcal{W} \to 0$ and $\psi \to 1$ at infinity (with respect to the compactly supported sources), and are regular at the $r = 0$ axis. In particular, regularity of the geometry requires $\psi$ to be an even function with respect to $r = 0$, and $\mathcal{W}$ needs to be an odd function. Furthermore, once the function $\psi$ has been determined, we can invert the relation and obtain the physical densities $\rho_{\text{D}}$ and $v^{\phi}$ to which these solutions correspond. As a general rule, a constant density rescaled source $\rho_{\text{N}}$ will lead to a real profile $\rho_{\text{D}}$ which is more dense towards the border. Roughly speaking, the effect of $\psi$, which is monotonically increasing function from infinity to the centre of the configuration, in Eqs.~\eqref{Eq:rhoNdef}--\eqref{Eq:vNdef} is to deplete the interior, i.e., to make it more dilute in the interior than the border, even if originally $\rho_{\text{N}}$ was a constant times a characteristic function. Still, these configurations are interesting enough for our purposes as we can easily find configurations exhibiting horizons. We choose units in which the volume of the region over which $\rho_{\text{N}}$ is supported is equal to $4 \pi /3$.

Not every choice of rescaled sources corresponds to physically acceptable matter, so after solving the constraints we recover $\rho_{\text{D}}$ and $v^{\phi}$ and check them. If the denominator in Eq.~\eqref{Eq:rhoNdef} becomes negative, i.e., $(v^{\phi})^2 \psi^4 r^2 > 1$, the dust moves superluminally and inverting the relation gives $\rho_{\text{D}} < 0$, violating the weak energy condition. We therefore retain only configurations with $\rho_{\text{D}} \geq 0$ and $(v^{\phi})^2 \psi^4 r^2 < 1$ everywhere, to numerical accuracy, and discard the rest.

\subsection{Asymptotic expansion and relation to asymptotic quantities}
\label{Subsec:Asymptotics}

In this subsection we present the asymptotic expansion of both functions defining the geometry, $\mathcal{W}$ and $\psi$, in terms of the multipole moments of the sources that we specify: $\rho_{\text{N}}$ and $v_{\text{N}}^{\phi}$. The purpose of doing it is twofold. First of all, the ADM mass $M_{\text{ADM}}$ and the angular momentum\footnote{We are considering an axisymmetric setup, hence here $J_{\text{ADM}}$ corresponds to the projection of the angular momentum on the direction orthogonal to the constant $z$ planes, i.e., it corresponds to rotations in the direction of $\partial_{\phi}$.} $J_{\text{ADM}}$ can be straightforwardly determined in terms of these multipole moments. Second, these expansions can be used to impose suitable boundary conditions for the numerical resolution of the equation in the outer boundary. 

\subsubsection{Multipole expansions of \texorpdfstring{$\psi$}{psi} and \texorpdfstring{$\mathcal{W}$}{W}}

We turn our attention to the multipole expression for $\mathcal{W}$. Starting from Eq.~\eqref{Eq:WPRescequation}, we denote the right hand side by $S(r,z) := 8 \pi r \rho_{\text{N}} v_{\text{N}}^{\phi}$. Furthermore, we find it convenient to multiply both sides of the equation by $e^{i \phi}$ which transforms Eq.~\eqref{Eq:WPRescequation} into
\begin{align}
    \tilde{\Delta}_{3D} \tilde{\mathcal{W}} (r,z,\varphi) := \tilde{S}(r,z,\phi)
    \label{Eq:AuxiliaryLaplacian},
\end{align}
where $\tilde{\Delta}_{3D}$ represents the flat space Laplacian and we have introduced the following functions: $\tilde{\mathcal{W}} (r,z,\phi) := \mathcal{W} (r,z) e^{i \phi}$ and $\tilde{S} (r,z,\phi) := S(r,z) e^{i \phi}$. We find convenient to change to spherical coordinates, $(r,z) \to (R,\theta)$, related with the cylindrical ones through the standard relation $r = R \sin \theta$ and $z = R \cos \theta$. In terms of them, the flat space line element reads
\begin{align}
    \dd s^2 = \dd R^2 + R^2 \left( \dd \theta^2 + \sin^2 \theta \dd \phi^2 \right). 
    \label{Eq:SphericalCoords}
\end{align}
From Eq~\eqref{Eq:AuxiliaryLaplacian}, we can immediately write down the multipole expansion of $\tilde{\mathcal{W}}$ in terms of the multipole moments of $\tilde{S}$~\cite{Jackson1998}:
\begin{align}
    \tilde{\mathcal{W}} = \sum_{\ell = 0}^{\infty} \sum_{m = - \ell}^{m = \ell} Q_{\ell m} \frac{Y_{\ell m} (\theta, \phi)}{R^{\ell + 1}}
\end{align}
where $Y_{\ell m} (\theta, \phi)$ represent the spherical harmonics (we stick to the Condon-Shortley convention for their phase) and the multipole moments $Q_{\ell m}$ are given by:
\begin{align}
    Q_{\ell m} := -\frac{1}{2 \ell + 1} \int_{M} \dd R \, \dd \theta  \, \dd \phi \,  \sin \theta R^{\ell+2} Y_{\ell m}^{\ast} (\theta, \phi) \tilde{S} (R,\theta,\phi). 
\end{align}
where $M$ represents the region where the source is supported. Plugging in our source, we automatically reach the conclusion that the only nonvanishing terms are those with $ m = 1$, and hence the leading order of the expansion is precisely the $Q_{11}$, whose explicit expression is:
\begin{align}
    Q_{11} := \sqrt{\frac{\pi}{6}} \int_{M} \dd \theta \, \dd R \, R^{3} \sin^2 \theta S(R,\theta),
\end{align}
\begin{align}
    \mathcal{W} =   \frac{\mathcal{W}_{\infty} \sin \theta}{R^2} + \order{\frac{1}{R^3}},
    \label{Eq:AsymptW}
\end{align}
with
\begin{align}
    \mathcal{W}_\infty = \frac{1}{4}  \int_{M} \dd \theta \, \dd R \, R^3 \sin^2 \theta S(r,z) = 2 \pi \int_{M} \dd r \, \dd z \, r^3 \rho_{\text{N}} (r,z) v_{\text{N}} (r,z),
    \label{Eq:AsymptWinfty}
\end{align}
where in the last equation we have transformed back to cylindrical coordinates.

We can repeat the same exercise for $\psi$, which in this case is simpler as the differential operator arising in the equation is already the Laplacian. In fact, the expansion for $\psi$ only contains terms with $m = 0$ due to axial symmetry and we get:
\begin{align}
    \psi(R,\theta) =1+\sum_{\ell = 0}^{\infty}\psi_{\ell 0} \frac{Y_{\ell 0} (\theta, \phi)}{R^{\ell + 1}},
\end{align}
where we notice that in this case the spherical harmonics reduce to simply Legendre polynomials $Y_{\ell 0} = \sqrt{\frac{2 \ell + 1}{4 \pi}} P_{\ell} \left( \cos \theta \right)$.
Since the leading behaviour of $\mathcal{W}$ is of the form $\mathcal{W} \sim R^{-2}$, we can perform a multipole expansion for $\psi$ as $R \to \infty$:
\begin{align}
    \psi (R,\theta) = 1 + \frac{M}{2R} + \order{\frac{1}{R^2}}, 
    \label{Eq:AsymptPsi}
\end{align}
where the factor of $2$ has been introduced for convenience and we have that $M$ is given by
\begin{align}
    M = 2 \pi \int_{M} \dd r \, \dd z \, r \rho_{\text{N}} (r,z) \, + \, \frac{1}{4} \int_{\mathbb{R}^3} \dd r \, \dd z \, r \psi^{-7} \left[   \left( \partial_r  \mathcal{W} \right)^2 + \frac{ \mathcal{W}^2}{r^2} - 2 \frac{ \mathcal{W} \partial_r  \mathcal{W}}{r} +  (\partial_z  \mathcal{W})^2 \right], 
    \label{Eq:AsymptM}
\end{align}
where we have again expressed the integral in cylindrical coordinates. In the case in which there is no rotation, the second integral identically vanishes. We notice that even though it has to be evaluated over the whole space, it decays as $1/R^2$ at infinity and hence we can estimate it numerically if we push the boundary to sufficiently large values.

\subsubsection{Determining the ADM quantities}
The spacetimes that we are considering are asymptotically flat so the ADM mass is well-defined and it only depends on the induced metric on a given Cauchy surface. For the slice for which we are solving the constraints, we can express the ADM mass as:
\begin{align}
    M_{\text{ADM}} = \frac{1}{16 \pi} \lim_{R \to \infty} \oint_{\mathcal{S}_{R}} \dd^2 x \sqrt{h} \sum_{j} \left( \partial_{j} \gamma_{ij} - \partial_{i} \gamma_{jj}  \right) \eta^{i},
    \label{Eq:MassADM}
\end{align}
where $\mathcal{S}_{R}$ represents a family of constant $R$ spheres, $\eta^i$ is their normal vector, $\dd^2 x$ represents the coordinates on the sphere and $\sqrt{h}$ the determinant of the induced metric on the sphere $\mathcal{S}_{R}$. Plugging the expression for the conformally flat metric, we find:
\begin{align}
    M_{\text{ADM}} = - \frac{1}{2 \pi} \lim_{R \to \infty} \oint_{\mathcal{S}_{R}} \dd^2 x \sqrt{h}\partial_i \psi \eta^i.
\end{align}
By plugging the asymptotic expression for $\psi$ obtained in Eq.~\eqref{Eq:AsymptPsi}, we find that with our conventions $M_{\text{ADM}}  = M$, i.e., it is purely determined in terms of its monopole moment. 

If the spacetime is also axisymmetric as in our case, it is possible to define an asymptotic ADM angular momentum which is aligned with the direction orthogonal to the subspaces generated by the orbits of the Killing field $\partial_{\phi}$ on a Cauchy slice as:
\begin{align}
    J_{\text{ADM}} = \frac{1}{8\pi} \lim_{R \to \infty} \oint_{\mathcal{S}_{R}} \dd^2 x \sqrt{h} \left( K_{ij} - K \gamma_{ij}\right) \eta^i \left( \partial_{\phi} \right)^{j}.  
\end{align}
We note that in this case $ K = 0$ and $K_{ij}$ can be fully expressed in terms of $\mathcal{W}$. In fact, it turns out that the asymptotic value of $\mathcal{W}_{\infty}$ as introduced in Eq.~\eqref{Eq:AsymptW} fully determines this angular momentum:
\begin{align}
    J_{\text{ADM}} = \mathcal{W}_{\infty},
\end{align}
which can be obtained in terms of the dipole moment of the source of the momentum constraint as expressed in Eq.~\eqref{Eq:AsymptWinfty}. We can combine them to introduce the dimensionless spin
\begin{align}
    \chi_{\text{ADM}} = \frac{J_{\text{ADM}}}{M_{\text{ADM}}^2},
\end{align}
which we expect to be smaller than $1$ for black hole configurations, as we will discuss below. 
\subsubsection{Mixed boundary conditions}
From the asymptotic behaviour of $\psi$ and $\mathcal{W}$ we can derive expressions that can be used to fix the asymptotic boundary conditions. In fact, we notice that if we take the derivatives of $\psi$ and $\mathcal{W}$ with respect to $R$ at large radii we obtain:
\begin{align}
    & \partial_R \psi = - \frac{M}{2 R^2} + \order{\frac{1}{R^3}}, \\
    & \partial_R \mathcal{W} = - 2 \frac{\mathcal{W}_{\infty} \sin \theta }{R^3} + \order{\frac{1}{R^4}}. 
\end{align}
We can use this to find the following relations
\begin{align}
    & \partial_{R} \psi + \frac{1}{R} \left( \psi - 1 \right) = \order{\frac{1}{R^3}}, \\
    & \partial_R \mathcal{W} + 2 \frac{\mathcal{W}}{R} = \order{\frac{1}{R^4}} , 
\end{align}
which have corrections that decay faster than $R^{-2}$ and $R^{-3}$. Hence, we can truncate the series to find suitable boundary conditions at infinity, more on this below. 

\subsection{Numerical solution of the constraints}
\label{Subsec:NumericalImplementation} 
We discretize the PDEs~\eqref{Eq:PsiRescequation}--\eqref{Eq:WPRescequation} over a 2D finite domain lattice. We choose fixed $\Delta z$ and $\Delta r$ grid spacings with boundary conditions at $z=0$, $z=L$, $r=0$ and $r=L$ for some choice of outer boundary distance $L$. Unless stated otherwise, assume that $\Delta z = \Delta r$. The numerical solving of the PDEs has been carried out via a parabolic diffusion-like relaxation scheme. 

Both elliptic PDEs are promoted to parabolic diffusion-like PDEs by adding a dependence of $\psi$ and $\mathcal{W}$ on the pseudo-time $t$, and adding the terms $-\partial_t \psi$ and $-\partial_t \mathcal{W}$ on the left-hand side of Eqs.~\eqref{Eq:PsiRescequation}--\eqref{Eq:WPRescequation}, i.e. we obtain the following parabolic PDE's,
\begin{align}
\label{eq:pseudo-time-psi}
    \partial_t \psi(r,z) &= \tilde\Delta \psi +  \frac{1}{4} \psi^{-7} \left[   \left( \partial_r  \mathcal{W} \right)^2 + \frac{ \mathcal{W}^2}{r^2} - 2 \frac{ \mathcal{W} \partial_r  \mathcal{W}}{r} +  (\partial_z  \mathcal{W})^2 \right] + 2 \pi \rho_{\text{N}},
    \\ \label{eq:pseudo-time-W}
    \partial_t \mathcal{W}(r,z) &= \tilde\Delta \mathcal{W} -\frac{\mathcal{W}}{r^2} -8\pi r \rho_{\text{N}} v^\phi_{\text{N}},
\end{align}
whose principal part is $\partial_t - \tilde{\Delta}$, where $\tilde\Delta$ is the flat space Laplacian operator. 
The PDE's are evolved in the $t$ coordinate specifying suitable initial data that we discuss below until a steady state solution $\partial_t \psi = 0$ and $\partial_t \mathcal{W}=0$ is reached, which corresponds to the solution of the original PDEs. We note that the PDE for $\mathcal{W}$ does not contain $\psi$, hence it can be solved in isolation if desired. Furthermore, it is linear and hence it can be solved through a Green function method, which requires performing a double integral. We have done it and compared it with the results obtained with the relaxation method and found a good agreement between both. In practice, we let both PDEs evolve together in parallel in the pseudo-time $t$. 

We are primarily interested in the stable end point of the diffusion-like PDEs, so we employ a first order accurate time integration scheme, namely the explicit Euler method. The particular trajectory taken towards the steady state solution is not important, provided that the evolution remains stable and converges to the unique steady state solution.

The spatial derivatives are evaluated using fourth order accurate, symmetric finite difference stencils. Second-order accurate stencils were also tested and found to produce qualitatively similar results. The resulting scheme therefore constitutes a first order accurate in time and fourth order accurate in space, explicit method-of-lines evolution.

For explicit Euler time integration of diffusion-like equations, the time step is constrained by a parabolic stability condition of the form
\begin{align}
    \Delta t \leq \alpha \Delta x^2,
\end{align}
where $\alpha$ is usually called the Courant-Friedrichs-Lewy (CFL) factor~\cite{Press2007}. For our equations, that are solved in cylindrical coordinates, we have found that the conservative value
\begin{align}
    \Delta t = 0.125 \,\Delta r\, \Delta z,
\end{align}
provides a robust and stable evolution.

The boundary conditions for $\psi$ at $z=0$ and $r=0$ are symmetric, correctly imposing the reflection symmetry in the $z=0$ plane and the rotational symmetry. $\mathcal{W}$ is symmetric about $z=0$ and antisymmetric about $r=0$. The outer boundaries use a discretised form of the mixed boundary conditions derived earlier truncating the asymptotic expansion to subleading order, 
\begin{align}
\label{eq:bdry_dirichlet_psi}
    \partial_R \psi + \frac{1}{R}(\psi-1) = 0,\\
\label{eq:bdry_dirichlet_W}
    \partial_R \mathcal{W} +2 \frac{\mathcal{W}}{R}  = 0 .
\end{align}
These boundary conditions are implemented at the numerical boundaries discussed above using a centred finite-difference approximation with the help of some additional ghost cells that are introduced for this purpose. 

The relaxation scheme requires us to specify a seed solution to initiate the pseudo-time evolution. The closer this seed is to the final solution, the faster the algorithm is expected to converge. In our case, instead of choosing the simplest choice, which is the vacuum metric, $\psi=1$ and $\mathcal{W}=0$, we employ the following initial seed for $\psi$ while still taking $\mathcal{W}=0$, which converges faster:
\begin{equation} 
\label{Eq:initial_seed}
    \psi_{t=0} =
        \begin{cases}
            1 + \dfrac{M}{2R}, & R > 1, \\[6pt]
            1 + \dfrac{M}{2}, & R \leq 1.
        \end{cases}
\end{equation}
It corresponds to the Schwarzschild metric in isotropic coordinates outside the coordinate unit radius sphere, while replacing the interior with a flat cap. The mass parameter $M$ is calculated from the integral of the energy density given in Eq.~\eqref{Eq:AsymptM}. We notice that this solution for $\psi$ would be exact outside the source in the absence of rotation and for a spherically symmetric source, hence it is clearly closer to the final solution than $\psi = 1$, even in the nonspherical cases. 
\subsubsection{Multigrid full approximation storage}
So far we have discussed the entire relaxation scheme on one single numerical grid. The final ingredient of our numerical solver is the adoption of the \emph{multigrid} method with \emph{full approximation storage}~\cite{Press2007}. 

\begin{figure}[htbp]
    \centering
    \includegraphics[width=0.6\textwidth]{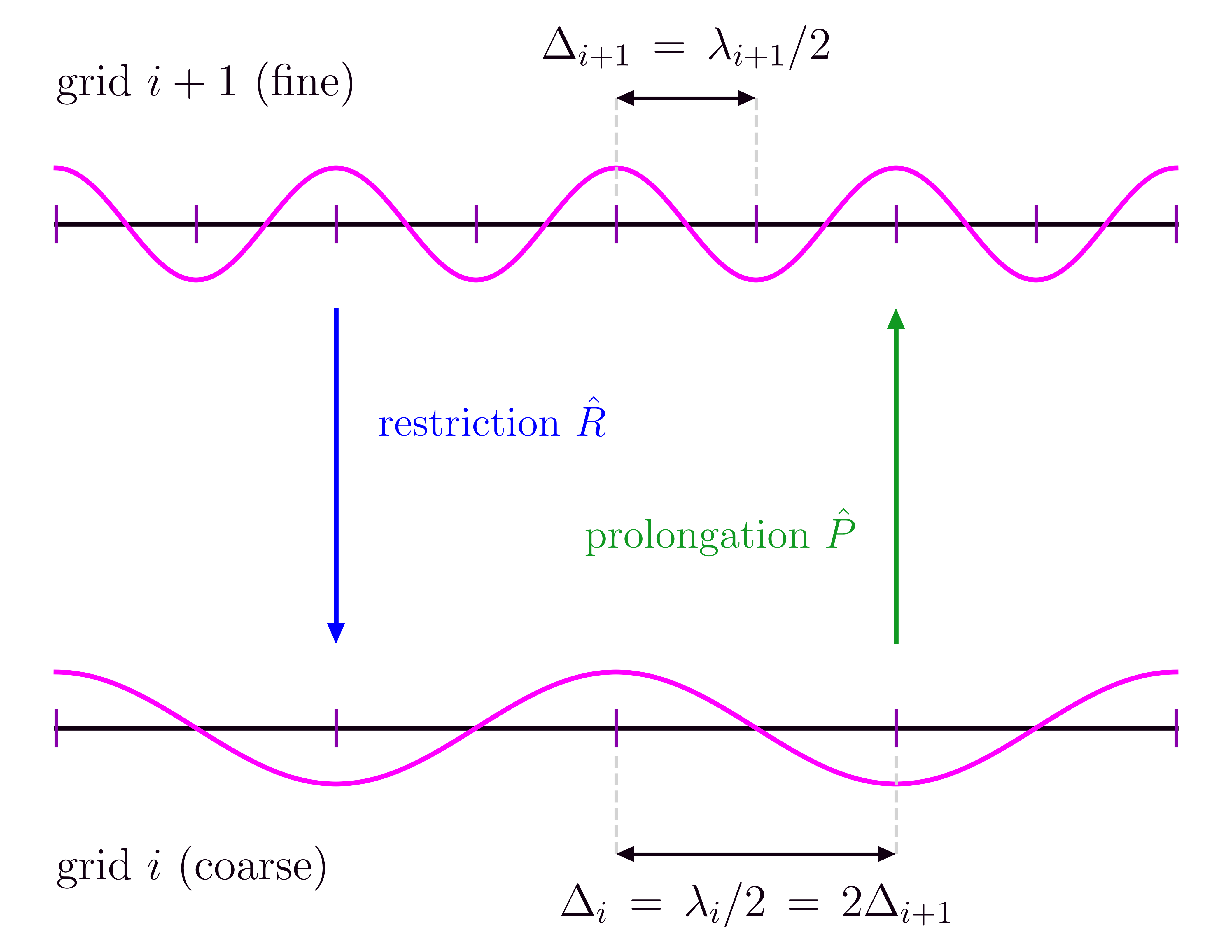}
    \caption{A pictorial representation of the multigrid algorithm that we use. It illustrates how the shortest wavelength that can be resolved by the $i$-th grid is twice the length of the spacing, namely $\lambda_i = 2 \Delta_i$. Hence each grid is best suited to removing errors of wavelength order $\lambda_i$. Furthermore, we represent the restriction and prolongation operators that allow to push the data from one grid to the other.}
    \label{fig:multigrid}
\end{figure}

The multigrid introduces a set of numerical grids, all covering the same domain. The base grid (grid $0$) has a grid spacing of $\Delta_0$ and each subsequent grid number $i$ has a spacing of $\Delta_i = \Delta_0 \cdot 2^{-i}$; i.e. halving the spacing each grid. Hence the time step evolution for each grid can be at most $\Delta t_i = \alpha \Delta^2_i = \alpha \Delta^2_0 \cdot 2^{-2i}$, to fulfil the stability criterion for parabolic equations. The coarse grid takes advantage of the largest time step which scales quadratically with grid spacing, and hence relaxes with fewer time steps. Relaxation schemes damp errors of wavelength $\lambda$ at a rate which scales as $(\Delta_i / \lambda)^2$; hence it is efficient to let coarser grids (with a larger $\Delta_i$) take care of the large scale structure of the solution while higher grids provide overall smoothing and finer detail resolution.
The power of the multigrid solver comes from the fact that each grid is best suited to removing errors with a wavelength $\lambda_i \approx 2 \Delta_i$, as pictorially displayed in Fig.~\ref{fig:multigrid}. 

The full approximation storage requires a small rewriting of the PDEs from Eqs.~\eqref{eq:pseudo-time-psi} and \eqref{eq:pseudo-time-W}; specifically, both are recast to
\begin{align} \label{eq:fas-pde}
    \partial_t \phi = \xi(\phi) - \tau
\end{align}
where $\phi $ can either be $\psi$ or $\mathcal{W}$. The \emph{residual} $\xi$ (of $\phi$) is equal to the original elliptic PDE of $\phi$, and is dynamically driven to $\tau$ which is an offset function that depends on all the variables of the system, namely the functions we are solving for, the spacetime coordinates and the sources. Its vanishing means we are identically solving the original PDEs. Since each grid will report a slightly different residual $(\xi)$ due to discretisation errors, $\tau$ represents this offset from the ``true residual'' that should be measured. We set $\tau=0$ for the finest grid, pretending it is the ``true residual''. Each grid below the finest calculates its offset $\tau_i$ (where $i$ refers to the grid number) as
\begin{align}
    \tau_i :=  \xi_i(\hat{R} \left[ \phi_{i+1} \right]) - \hat{R} \left[ \xi_{i+1}(\phi_{i+1})\right] ,
\end{align}
where $\hat{R}$ is the restriction operator, interpolating values from grid $i+1$ to $i$. Breaking down the meaning of these terms, $\hat{R}\left[\xi_{i+1}(\phi_{i+1})\right]$ calculates the residual $\xi_{i+1}(\phi_{i+1})$ on the finer grid and restricts that down to the coarse grid. Instead, $\xi_i(\hat{R} \left[ \phi_{i+1})\right]$ means to first restrict the field $\phi_{i+1}$ onto the coarse grid, and then use the evaluation of $\xi_i$ at the coarse grid of the PDE residual. The difference of these is the estimate of the bias $\tau$ that we need to compensate for the discretisation error in calculating $\xi$ on coarser grids. This ensures that all grids converge toward the same solution, avoiding wasted computations caused by the grids pulling each other’s solutions toward different attractors.

Putting everything together, we start by giving the initial data seed from Eq.~\eqref{Eq:initial_seed} to each grid and perform a set of \emph{V-cycles}; typically a few tens of these V-cycles reduce the residual to machine precision. The V-cycle begins by repeating a set of downwards directed steps between a fine and a coarse grid, starting from the finest and finishing on the coarsest grid. Each downward step does $\nu_2 =3$ smoothing steps\footnote{We have also used $\nu_2 = 2$ and no significant difference is observed.} evolving Eq.~\eqref{eq:fas-pde} in pseudo time then sets $\tau$ on the grid below while also replacing $\hat{R} \left[\phi_{i+1} \right] = \phi_i$. Once the coarsest grid is reached, with $\tau_0$ we perform $\nu_1 = 512$ relaxation steps, this is the bottom of the V-cycle. Since the coarse grid has the smallest number of points, these $\nu_1$ steps are negligibly fast. Next a set of upwards steps from a coarse to a fine grid are repeated until the finest grid is reached. Specifically we use
\begin{align}
    \phi_i \rightarrow \phi_i + \hat{P} \left[ \phi_{i-1} - \hat{R} \left[ \phi_i \right] \right] 
\end{align}
where $\hat{P}$ is the prolongation operator, interpolating values from grid $i-1$ to $i$. After prolongation, again $\nu_2$ smoothing steps are done. Note that $\phi_{i-1}$ here has undergone some relaxation steps, with the offset $\tau_{i-1}$ as well as corrections from lower grid prolongation steps. This process is repeated until the relative residual on the finest grid defined as
\begin{align}
    \mathcal{E} \;=\; \sqrt{\frac{\lVert \xi_\psi \rVert_{2}^2 + \lVert \xi_{\mathcal{W}} \rVert_{2}^2}{\lVert s_{\psi}\rVert_{2}^2 + \lVert s_{\mathcal{W}}\rVert_{2}^2} },
\end{align}
satisfies $ \mathcal{E} \;<\;  10^{-9}$, at which point the solution is declared converged. Here $s_\psi$ and $s_{\mathcal{W}}$ are the matter sources defined as,
\begin{align}
\label{Eq:source_psi}
     |s_{\psi}| &= 2 \pi \rho_{\text{N}}, \\
    \label{Eq:source_W}
     |s_{\mathcal{W}}| &= 8 \pi r\rho_{\text{N}} v^\phi_{\text{N}} ,
\end{align}
and $\lVert f \rVert_{2}$ represents the $L^{2}$ norm over the axisymmetric domain,
\begin{align}
    \lVert  f \rVert_{2} = \left(\, \int |f|^2 \, r \,\dd r \, \dd z  \,\right)^{\frac{1}{2}}\,.
\end{align}
Finally we note that $s_{\psi}$ does not vanish for us since we never consider vacuum cases, but in vacuum it could be exchanged with a constant. A detailed convergence analysis of the algorithm is reported in App.~\ref{App:Convergence}.

\section{Apparent horizons}
\label{Sec:Horizons}

We present the equation that an apparent horizon needs to satisfy in Subsec.~\ref{Subsec:ApparentHorizons}, the properties that we can associate with it, namely the mass and angular momentum, in Subsec.~\ref{Subsec:MassAngularMomentum} and how we find them numerically in Subsec.~\ref{Subsec:NumericalHorizon}.

\subsection{Apparent horizons: MOTS}
\label{Subsec:ApparentHorizons}

Apparent horizons are defined as the outermost marginally outer trapped surface (MOTS) on a given spatial hypersurface. This means that they are closed two-dimensional surfaces such that the expansion of outgoing null geodesics orthogonal to them is zero everywhere on the surface~\cite{Alcubierre2008}. Consider a surface with unit normal vector $s^\mu$ embedded in one of the Cauchy slices that we considered in Sec.~\ref{Sec:InitialData}. With the help of the normal vector to the Cauchy slices that we have chosen, $n^{\mu}$, introduced in Eq.~\eqref{Eq:n_vector}, we can build the future directed null normals as $V^\mu_{\pm} = n^{\mu} \pm s^{\mu}$. The expansion of both congruences are given by
\begin{align}
    \Theta_{\pm} = K_{ij} s^i s^j - K  \pm D_i s^i. 
\end{align}
We note that the last term is actually the mean curvature of the closed surface that we are considering. To see this, we can introduce the two-dimensional metric induced on the surface $h_{ij} := \gamma_{ij} - s_i s_j$ and its extrinsic curvature as:
\begin{align}
    k_{AB} = P_{A}^{\ i} P_{B}^{\ j} D_i s_j, 
\end{align}
where $P_{A}^{\ i} = h_A^{\ i}$ is the projector onto the surface. The mean curvature is then simply its trace $k := D_i s^i$. 

Axisymmetry implies that the apparent horizon, in case it exists, should also be axisymmetric. Hence, we find convenient to switch again to spherical coordinates as introduced above Eq.~\eqref{Eq:SphericalCoords}. In these coordinates, the horizon would be described by a function of the polar angle alone, i.e., an equation of the form $R = f (\theta)$. The unit normal vector $s^i$ in these coordinates will be given by
\begin{align}
    s^{i} = \frac{\psi^{-2}}{\sqrt{R^2+ \left(f' \right)^2}} \left( R \delta^{i}_{\ R} - \frac{f^{\prime}}{R} \delta^i_{\ \theta} \right).
\end{align}
Notice that in our case we have $K = 0$. Hence, the nonvanishing components of $K_{ij}$ are only the nonvanishing components of $\left( \boldsymbol{L} W \right)_{ij}$ and the contraction with $s^i s^j$ vanishes. Thus, in our setup the expansions reduce to:
\begin{align}
    \Theta_{\pm} = \pm k.    
\end{align}
Hence, in this case, the MOTS correspond to minimal surfaces, i.e., those with zero mean curvature $k = 0$. We can evaluate the curvature explicitly:
\begin{align} \label{Eq:div_s}
    k = D_i s^i = \frac{1}{\sqrt{\gamma}} \partial_i\big( \sqrt{\gamma} \,s^i \big) = 0, \quad \text{with} \quad \sqrt{\gamma} = \psi^6 R^2 \sin(\theta),
\end{align}
and setting it to zero we obtain the ODE that the horizon should satisfy:
\begin{align}
    R'' + \left( \frac{R^{\prime 3}}{R^2} + R' \right) \left( 4 \frac{\partial_{\theta} \psi}{\psi} + \cot \theta \right) - R^{\prime 2} \left( \frac{3}{R} + 4 \frac{\partial_R \psi}{\psi} \right) - 2 R - 4 R^2 \frac{\partial_{R} \psi}{\psi} = 0. 
    \label{Eq:Horizon}
\end{align}
We are representing the surface of the horizon as $R(\theta)$ to avoid a more cumbersome notation and the conformal factor $\psi(r,z)$ obtained in cylindrical coordinates should be evaluated on $\psi \left( R(\theta) \sin \theta,  R(\theta) \cos \theta \right)$. We note that our equation agrees with Eq.~(37) of Ref.~\cite{Eppley1977} when taking $q = 1$, as the framework there is different from ours: there the extrinsic curvature vanishes (here it just does not contribute) and the metric is conformally flat only when $ q = 1$. We note that it should also agree with the one given in Eq.~(3.14) of Ref.~\cite{Nakamura1988} but it contains a typo: instead of a derivative with respect to $R$, the last term contains a derivative with respect to $\theta$. 

Even though the extrinsic curvature drops from the equations, the equation for $\psi$ still contains a term related to the rotation that is acting as an additional source and consequently the existence and location of an apparent horizon still depends on the rotation. Furthermore, we note that we focus on data that are reflection symmetric with respect to the $z = 0$ plane, i.e., the solutions are symmetric under $z \to - z$ or, equivalently in spherical coordinates, $\theta \to \pi - \theta$. Then, we have that the horizon $R(\theta)$ must satisfy $R(\theta) = R (\pi - \theta)$ and necessarily $R '(\pi/2) = 0$. Furthermore, regularity of the surface (we recall that it is a closed surface) requires $R ' (0 ) = R' (\pi) = 0$. 

In general, a MOTS is not necessarily a horizon since it is possible that there are outwards deformations of it which do not change its character. A precise mathematical criterion for it being a horizon, namely that it is strictly stable under deformations within the Cauchy slice, was presented in Refs.~\cite{Andersson2005}. Roughly speaking, any small deformation of it should convert it into a spacelike surface to ensure that it is a horizon: it would be the surface that sharply separates the trapped from the antitrapped region. In our cases, we expect the MOTS found to be the apparent horizons.

\subsection{Apparent horizons: Hawking mass and angular momentum}
\label{Subsec:MassAngularMomentum}
Once we have found some initial data satisfying the GR constraints and an apparent horizon, i.e., a closed surface that satisfies Eq.~\eqref{Eq:Horizon} for a minimal surface, we can compute its area and angular momentum. First of all, we can straightforwardly compute its invariant area $\mathcal{A}_{\mathcal{H}}$ as
\begin{align}
    \mathcal{A}_{\mathcal{H}} = \oint_{\mathcal{H}} \dd^2 x \sqrt{h} = 2 \pi \int_0^{\pi} \dd \theta \sin \theta \,\psi^4 R^2 \sqrt{1+(R'/R)^2},
    \label{Eq:HorizonArea}
\end{align}
where $\sqrt{h}$ represents again the determinant of the induced metric on the horizon surface, and the conformal factor $\psi$ should be evaluated on the horizon $\mathcal{H}$. We can compute the irreducible mass associated with the horizon in terms of its area through the following relation
\begin{align}
    M_{\mathcal{H}} = \sqrt{\frac{\mathcal{A}_{\mathcal{H}}}{16 \pi}}. 
    \label{Eq:HawkingMass}
\end{align}
We notice that this agrees with the Hawking mass of the apparent horizon as originally introduced in Ref.~\cite{Hawking1968}, since the expansion contribution vanishes for horizons. Furthermore, we can associate an angular momentum with any axisymmetric horizon (see, e.g.,~\cite{Jaramillo2010} and references therein) through the expression:
\begin{align}
    J_{\mathcal{H}} = \frac{1}{8\pi}  \oint_{\mathcal{H}} \dd^2 x \sqrt{h} \left( K_{ij} - K \gamma_{ij}\right) \eta^i \left( \partial_{\phi} \right)^{j}.
\end{align}
We can explicitly express the integral in terms of $\mathcal{W}$ 
\begin{align}
    J_{\mathcal{H}} = \frac{1}{4} \int^\pi_0 \dd \theta R \sin\theta \Big[ R'(\mathcal{W} \cos \theta -\partial_\theta \mathcal{W} \sin \theta)+ R \sin \theta (R \partial_R \mathcal{W} -\mathcal{W})\Big],
    \label{Eq:HorizonAngMom}
\end{align}
where all the quantities should be evaluated on the horizon surface $R = R(\theta)$. We can combine it with $M_{\mathcal{H}}$ to define the total mass $M_T$ of the horizon applying Christodoulou's formula~\cite{Christodoulou:1970wf,Christodoulou1971}
\begin{align}
    M_{T}^2 = M_{\mathcal{H}}^2 + \frac{J_{\mathcal{H}}^2}{4 M_{\mathcal{H}}^2} = \frac{\mathcal{A}_{\mathcal{H}}}{16 \pi} + 4 \pi \frac{J_{\mathcal{H}}^2}{\mathcal{A}_{\mathcal{H}}}. 
    \label{Eq:TotalMass}
\end{align}
The reason for this definition is that in general the total mass of the black hole can decrease for rotating black holes. Processes like the Penrose process~\cite{Penrose1969,Penrose1971,Wald1974} allow the extraction of energy and angular momentum from the black hole. The second law of black hole dynamics states that the area of the black hole horizon itself should not decrease, which is proportional to the irreducible mass. Furthermore, it is also useful to introduce the dimensionless spin of the horizon as:
\begin{align}
    \chi_{\mathcal{H}} = \frac{J_{\mathcal{H}}}{M_{T}^2},
\end{align}
which for Kerr black holes should satisfy $\abs{\chi_{\mathcal{H}}} \leq 1$ to avoid the presence of naked singularities. 

Naively, if the entire matter distribution is inside the horizon, we would expect the ADM mass from Eq.~\eqref{Eq:MassADM} to agree with the total mass of the horizon $M_T$ introduced in Eq.~\eqref{Eq:TotalMass}. However, in our case, since the horizons will not in general enclose the full matter content, we expect $M_T \lesssim M_{\text{ADM}}$.

Something similar happens with the angular momentum. We notice that if the horizon encloses all the matter content, then a standard computation (following for instance~\cite{Poisson2009}) allows to show that the angular momentum computed at the horizon $J_{\mathcal{H}}$ and at infinity $J_{\text{ADM}}$ should agree. In general, in our case the horizon will intersect the matter content and hence we expect $J_{\text{ADM}} \gtrsim J_{\mathcal{H}}$ as an additional contribution from the matter region is expected to occur. 

\subsection{Numerical computation of the horizon}
\label{Subsec:NumericalHorizon}

To find the apparent horizon, we define a generic surface of revolution $\mathcal{F}=0$, 
\begin{align}
    \mathcal{F} = R - f(\theta),
\end{align}
where $f(\theta)=f(\pi - \theta)$ implies reflection symmetry about the $z=0$ plane.

Numerically, we discretise the surface over a uniform set of cell centred points on $0<\theta<\pi/2$. The set of discrete angles are $\theta_i = (i-1/2) \Delta \theta$, where $\Delta\theta = \pi/(2 N_\theta)$ and $i \in [1,2,...,N_\theta]$. In other words, the smallest angle is $\Delta \theta/2$ and the largest is $\pi/2 - \Delta \theta/2$.
This choice of coordinate samples avoids the coordinate singularity at the pole $\theta=0$.

Over the discrete angles, we define the set of spherical radii $f_i = f(\theta_i)$ which define the shape of the surface in discretised form. The boundary conditions for $f_i$, as well as any other data on the surface, are both symmetric at the pole and the equator ($\theta = \pi/2$). We have used three different methods, a shooting and two relaxation schemes applied to two variants of the zero-expansion equation. The three of them seem to provide quite similar results.

\subsubsection{Shooting method}

The method works as follows. One makes an initial guess for the radius of the horizon $R_{\ast}$ at the closest value of $\theta$ to the equator which we can achieve with our grid, namely \mbox{$\theta = \pi/2 - \Delta \theta/2$}. It should be outside the MOTS. In practice, this can be easily done by making the rough approximation of where the horizon would be in case the solution was spherically symmetric with the calculated ADM mass (the solution would be Schwarzschild) and simply taking $R_{\ast}$ to be an order of magnitude bigger. In App.~\ref{App:HorizonAppendix} we have derived the Taylor expansions for the equation of the horizon, namely Eq.~\eqref{Eq:Horizon}. The resulting Eqs.~\eqref{Eq:EquatorRShoot}--\eqref{Eq:EquatorRprimeShoot} can be used to derive approximate initial conditions for the horizon equation by truncating the expansions to leading order: 
\begin{align}
    & R(\theta_{\ast} ) = R_{\ast}, \\
    & R' (\theta_{\ast}) = - \Delta \theta \left( R_{\ast} + 2 R_{\ast}^2 \partial_R \log \psi (R_{\ast}, \pi/2 - \Delta \theta /2 )  \right).
\end{align}
The horizon equation is integrated along decreasing values of $\theta$, i.e., towards the pole, until the minimum angle of the grid is hit \mbox{$\theta_{\text{f}} = \Delta \theta/2$.} At this angle, if the surface is really the horizon, we expect to approximately reach the values:
\begin{align}
    & R(\theta_{\text{f}} ) = R_{\text{f}}, \\
    & R' (\theta_{\text{f}}) =  \frac{\Delta \theta}{2} \left( R_{\text{f}} + 2 R_{\text{f}}^2 \partial_R \log \psi (R_{\text{f}},\Delta \theta /2 )  \right), \label{Eq:PoleDerivative}
\end{align}
where again the Taylor series in Eqs.~\eqref{Eq:PoleRShoot}--\eqref{Eq:PoleRprimeShoot} have been truncated to leading order. If the value of $R' (\theta_{\text{f}})$ is too large, the surface has not closed enough, meaning that the initial guess $R_{\ast}$ was too large. Conversely, if $R^{\prime}_{\text{f}}$ is below the expected value in Eq.~\eqref{Eq:PoleDerivative}, it means that the original guess $R_{\ast}$ was too small. By varying the ansatz with this information, one can get a better approximation towards the horizon surface. In case that there is no horizon, one will simply reach $R_{\ast} = 0$ while decreasing the value of the initial guess. 

Hence, in summary, the overall procedure goes as follows. One starts with a guess $R_{\ast}$ which is larger than the expected value for the horizon, as discussed above, and integrates towards the pole. This procedure is repeated until the value of the derivative at the pole is below the one expected. This initial guess for which the value of the derivative is below the expected result, let us call it $R_{\ast \text{L}}$, and the previous value for which the derivative was still above, $R_{\ast \text{U}}$, are used to carry on a bisection search to narrow the gap between the two until machine precision is reached. 

\subsubsection{Relaxation solvers}
The remaining two algorithms used to find the horizon are essentially both relaxation solvers taking advantage of pseudo-time parabolic diffusion, as discussed in Sec.~\ref{Subsec:NumericalImplementation}.  As they have a one-dimensional domain, it is fast enough to solve at the coarsest grid, and cascade the solution up to higher grids. We can resolve until the required precision is reached without the need for multigrid. They both enforce symmetric boundary conditions, namely $R'(0)=R'(\pi/2)=0$, redefining derivative stencils near the boundaries to enforce them.

The first method, we call relaxation, promotes Eq.\eqref{Eq:Horizon}, to a diffusion-like equation on $R(\theta)$ reading,
\begin{align} \label{Eq:relaxation_diffusion}
    \partial_t R(\theta) &:= \partial_\theta^2 R + Z(R, \psi, \theta, \partial_\theta R, \partial_\theta \psi, \partial_R \psi),
\end{align}
where $Z$ is sum of irrelevant nonlinear lower derivative terms. A single pseudo-time step has $\Delta t = \alpha (\Delta \theta)^2$ and a theoretical maximum CFL factor of $\alpha=0.5$, but we use the conservative choice $\alpha=0.25$.

The second relaxation type method, we denote relaxation-flow, is somewhat similar to the hyperbolic flow method presented in Ref.~\cite{Etienne:2025auk}, but the method we use is first order in time, which is a parabolic flow. The equation $\Theta^+=0$ is solved by evolving a surface in pseudo-time according to the ``flow equation'',
\begin{align} \label{Eq:flow_equation}
\partial_t R(\theta) := -\eta \Theta^+,    
\end{align}
where $\eta$ is a diffusion rate parameter we set to unity. 
Eq.\eqref{Eq:div_s} is used to calculate $\Theta^+(R(\theta),\theta)$ over the surface which transforms the flow equation to,
\begin{align}
    \partial_t R(\theta) = \frac{1}{\lambda(\theta) R^2} \partial_\theta^2R\;+ \;Y(R, \psi, \theta, \partial_\theta R, \partial_\theta \psi, \partial_R \psi),
\end{align}
where $Y$ contains irrelevant lower derivative terms. The function $\lambda = \psi^2 (1+(R'/R)^2)^{3/2}$ exhibits the property $\lambda(\theta)\geq 1$ since $\psi \geq 1$. Thus, the flow equation is the same as Eq.~\eqref{Eq:relaxation_diffusion} but with an extra factor of $1/(\lambda (\theta) R^2)$ multiplying the $\partial_\theta^2R$ terms and some unimportant differences in the other nonlinear terms $Y$ and $Z$. We take the time step to be $\Delta t = 0.25 R^2 \Delta \theta^2 $ which is a conservative choice, since the maximum theoretical stable time step is  $\Delta t = \alpha ( \lambda R^2 \Delta \theta^2)$ for a CFL factor of $\alpha=0.5$, and we recall that $\lambda$ is never smaller than one.

The advantage shared by both the relaxation solvers is their robustness to bad initial guesses and highly unusual geometries. As a proof of principle, we find that for a surface with small deviation $R = R_H + h(\theta,\phi)$ from the Schwarzschild horizon, using the decomposition of such a surface in spherical harmonics presented in App.~\ref{App:NearHorizon}, we find that the flow relaxes all the modes to zero, in particular they decay exponentially. The greater the value of $\ell$, the faster they decay.

\subsubsection{Comparison and convergence testing}
To check the convergence of the three different methods we have considered the following conformal factor:
\begin{align}
    \psi (R , \theta) = 1 + \frac{M}{2R} + \epsilon \left( \frac{M}{2R} \right)^3 P_{2} \left( \cos \theta \right),
\end{align}
where $\epsilon$ is taken to be a small parameter and we have computed to leading order in $\epsilon$ how the horizon changes with respect to the constant $R = M/2$ Schwarzschild counterpart: 
\begin{align} \label{eq:ecc_bh_pert}
    R(\theta) = \frac{M}{2} \left[ 1 + \epsilon \frac{5}{7} P_{2} \left( \cos \theta \right) \right] + \order{\epsilon^2}.
\end{align}
\begin{figure}[htbp]
    \centering
    \includegraphics[width=0.655\textwidth]{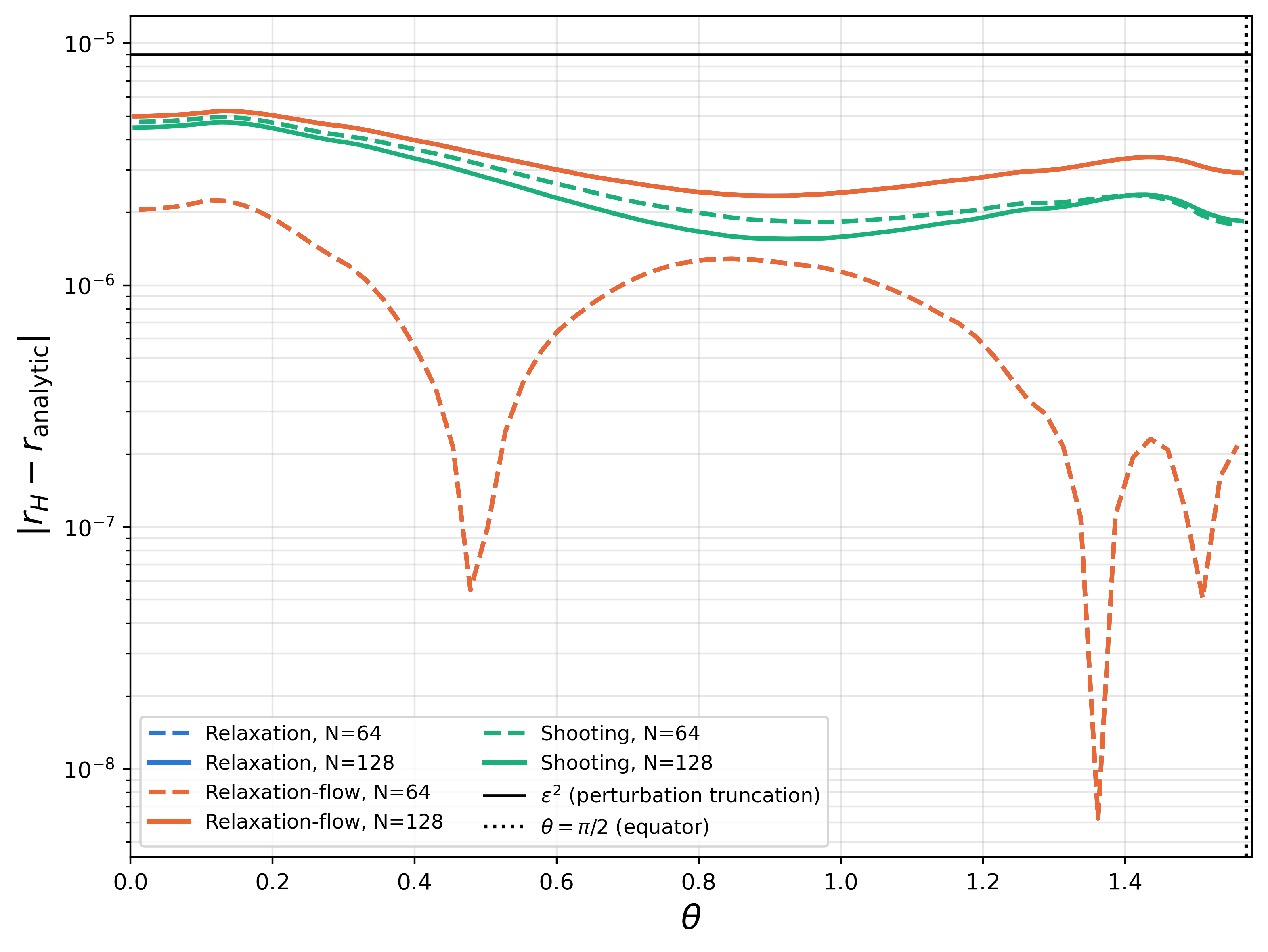}
    \caption{Absolute difference between the numerical apparent horizon $R(\theta)$ and the first-order analytic result, Eq.~\eqref{eq:ecc_bh_pert}, for the three horizon finders: relaxation (blue), relaxation-flow (red), and shooting (green). Dashed and solid lines correspond to $N=64$ and $N=128$ points on the surface, respectively. The relaxation curves lie exactly beneath the relaxation-flow ones and are therefore not visible. The solid black line estimates the size of the neglected $\mathcal{O}(\epsilon^2)$ term, which bounds the expected discrepancy.}
    \label{fig:ecc_bh_conv}
\end{figure}

In Fig.~\ref{fig:ecc_bh_conv}, the numerical solution of the three horizon finding methods are shown. We find that the deviation with respect to the analytic solution at order $\epsilon$ in Eq.~\eqref{eq:ecc_bh_pert} is smaller than $\epsilon^2$, hence we conclude that in all the cases the numerical solver agrees well with the analytic approximation computed. Additionally, Fig.~\ref{fig:ecc_bh_conv} shows that the small discrepancy between the shooter and the two relaxation based algorithms for different resolutions. Throughout this work we run the three horizon finder algorithms in parallel as a three way checker.

\section{Gravitational collapse and the hoop conjecture}
\label{Sec:Hoops}

In this section we review the hoop conjecture and we explain how it can be made mathematically precise in our setup. We discuss in Subsec.~\ref{Subsec:ReviewHoop} the original motivation for the conjecture together with the imprecision in its formulation, we discuss how it can be interpreted in our specific setup in Subsec.~\ref{Subsec:AxialSym}, and in Subsec.~\ref{Subsec:OverSpinning} we discuss the interplay between the hoop conjecture and naked singularities.

\subsection{The hoop conjecture}
\label{Subsec:ReviewHoop}

The original motivation of Thorne for introducing the hoop conjecture in~\cite{Thorne1972} was the study of a toy model consisting of the collapse of an infinitely elongated cylindrical distribution of matter. He showed that in such a situation, the end state of the configuration did not exhibit any trapped region and hence no black hole forms. In that sense, this motivated him to introduce the following conjecture: ``horizons form when and only when a mass $M$ gets compacted into a region whose circumference $\mathcal{C}$ in every direction satisfies $\mathcal{C} \lesssim  4 \pi GM \,$'', where the symbol $\lesssim$ intends to say that the conjecture should be understood up to an undetermined numerical constant.

In other words, it claims that a configuration will exhibit a horizon if and only if a hoop of circumference equal to the gravitational radius of a region to which a mass $M$ can be associated (up to an undetermined numerical constant), can be passed through all the directions of the configuration in such a way that the matter content lies inside of it. 

The conjecture is imprecise for several reasons. The first feature that stands out is the $\lesssim$ symbol, which intends that there is a multiplicative factor that is not specified. In fact, it has been proposed that different constants should hold for the if and the only if parts of the conjecture~\cite{Flanagan1991,Flanagan1992}.

The second problem is somewhat more subtle, as it stems from the absence of a local definition of mass and energy in GR: energy is not localised in GR. This issue is closely related to the difficulty of quantifying the energy of the gravitational field itself, which becomes particularly pronounced for vacuum configurations. There have been some proposals for a characterization of the mass inside a given bounded region of spacetime~\cite{Penrose1982,Brown1992,Liu2003}, where the most compelling one seems to be the one in Ref.~\cite{Wang2008}\footnote{This does not attempt to be an exhaustive list, see~\cite{Szabados2004} for a review.}. However, to the best of our knowledge none of them has been systematically studied as a candidate to be the mass entering the formulation of the hoop conjecture. 

Finally there is the subtlety associated with the notion of horizon itself. The notion of event horizon is a teleological concept and hence it is not sensible to formulate the conjecture in terms of it. The most common approach is to resort to some local definitions of horizon, for instance the formulation in terms of MOTS, as defined for instance in~\cite{Gourgoulhon2005}. In particular, the stability of the MOTS is required~\cite{Andersson2005} in order to ensure that they are separating the trapped and nontrapped regions. However, this notion is problematic since the presence of trapped surfaces depends on the slicing of the spacetime that one chooses, see for instance~\cite{Wald1991}, and such a slicing choice is required to define the circumference. 

There have been many attempts to reformulate the conjecture in a precise way as well as many attempts to prove or disprove it in specific setups where it was possible to make sense of it, see~\cite{Flanagan1991,Senovilla2007,Wald1997} for a non-exhaustive list. However, to the best of our knowledge, an explicit analysis of the impact of rotation on the conjecture has not been carried out, although Ref.~\cite{Senovilla2007} discusses its role. We present this work as a first attempt to fill in this gap in the literature. 

\subsection{Axially symmetric setups}
\label{Subsec:AxialSym}

\begin{figure}[htbp] 
    \centering
    \includegraphics[width=0.8\textwidth]{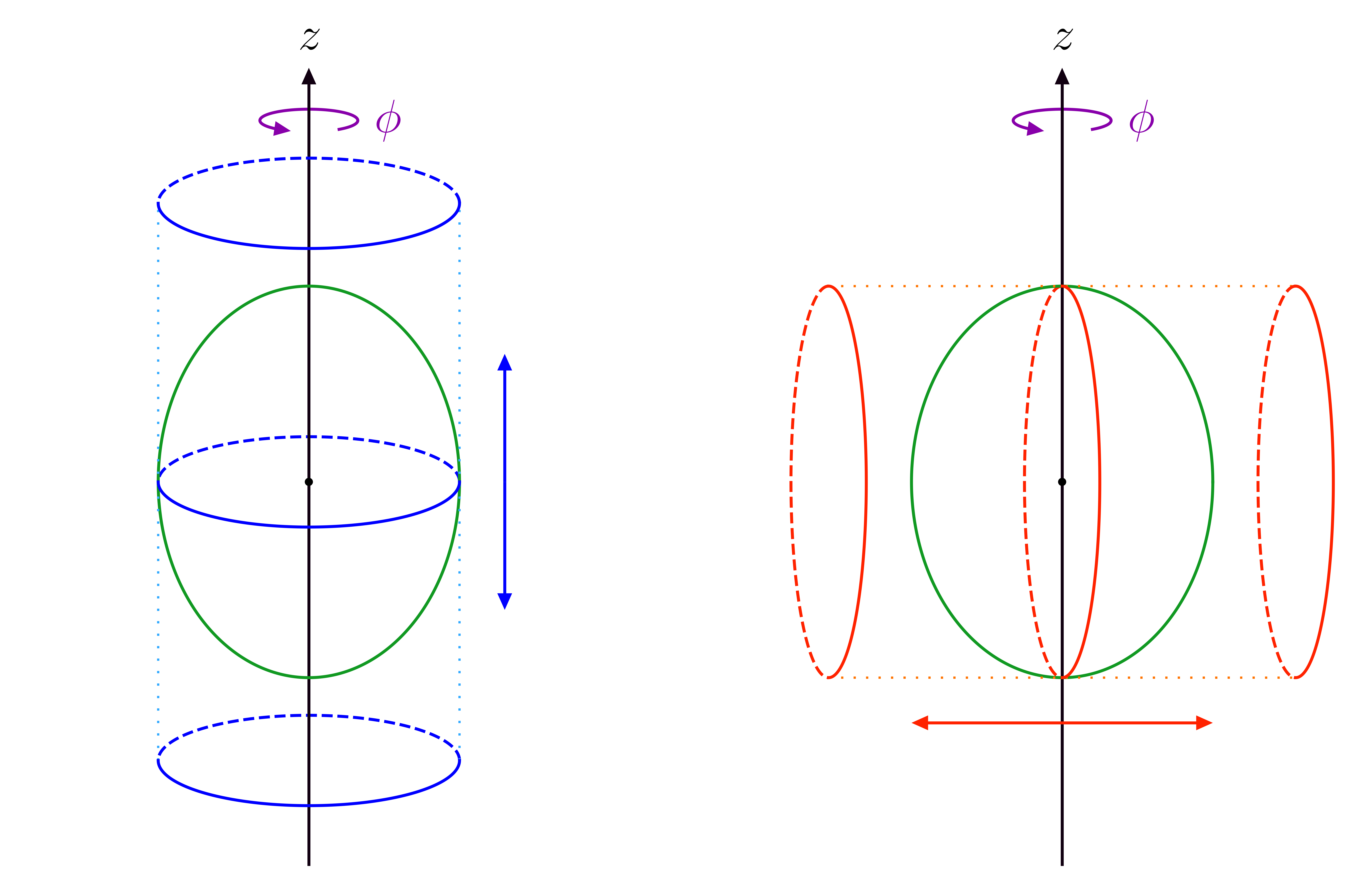}
    \caption{Example ellipsoidal surface of revolution in $\phi$ plotted in green. The blue hoop represents the equatorial circumference with $\theta = \pi/2$ and red the polar circumference with constant $\phi$.}
    \label{Fig:Circumfs}
\end{figure}

We now explain how the ambiguities in the formulation of the hoop conjecture can be overcome in the setup that we are discussing. To begin with, we consider how to measure the length of the relevant circumferences in this setting. The fact that the matter content is not radially infalling in this choice of coordinates singles out the time slicing that we are choosing. Hence, it is natural to compute the circumferences with respect to the induced metric on this 3D slicing. Due to axisymmetry, the relevant surfaces will also be axisymmetric, meaning that, like the horizon, they can be described in spherical coordinates $(R,\theta,\phi)$ simply by a function $ R = R(\theta)$ in such a way that the induced metric on the surface is given by:
\begin{align}
    \dd s^2 = \psi^4 R^2  \left[ \left( 1 + \left( \frac{R'}{R} \right)^2 \right) \dd \theta^2  + \sin^2 \theta \dd \phi^2 \right],
\end{align}
where we are taking the pullback to the surface of all of the functions above. For a given ellipsoidal surface embedded in the axially symmetric 3D space, we can introduce the polar and equatorial circumferences, which are pictorially represented in Fig.~\ref{Fig:Circumfs}. 
They correspond to the curves on the surface which have fixed azimuthal coordinate $\phi$ and fixed polar coordinate $\theta$, respectively. Both are geodesic curves over the 2D surface. In particular, the curves can be parametrized as $E \equiv (\theta = \pi/2, \phi )$ for $\phi \in (0, 2 \pi)$ and $P \equiv (\theta, \phi_0)$ for $\theta \in (0, \pi)$ and $\phi_0$ any constant value for the azimuthal angle. Their length is given by: 
\begin{align}
    & \mathcal{C}_E = 2 \pi R (\pi/2) \psi^2 \left[ R (\pi/2), \pi/2 \right] , \label{Eq:EquatorialCircumf} \\ 
    & \mathcal{C}_P = 2 \int_0^{\pi} \dd \theta R(\theta) \psi^2 \left[ R (\theta), \theta \right]  \sqrt{1 + \left( \frac{R' (\theta)}{R(\theta)} \right)^2}. \label{Eq:PolarCircumf}
\end{align}
We need to decide now which surfaces $R(\theta)$ are relevant for our purposes and how to compute the mass that they enclose. In general there are two natural candidates. 

First of all, if the geometry exhibits a horizon, that we can numerically compute with high precision as described in Sec.~\ref{Sec:Horizons}, we can compute the equatorial and polar circumferences on the horizon surface by directly applying Eqs.~\eqref{Eq:EquatorialCircumf}--\eqref{Eq:PolarCircumf} to the obtained surface, we will call them $\mathcal{C}^{\text{AH}}_E$ and $\mathcal{C}^{\text{AH}}_P$ respectively. There are two candidates for the mass enclosed by such a surface in the rotating case. We could consider either the Hawking (the irreducible) mass introduced in Eq.~\eqref{Eq:HawkingMass} or the total mass introduced in Eq.~\eqref{Eq:TotalMass}, which we recall includes the angular momentum contribution. Hence, we introduce the following ratios:
\begin{align}
    & \mathcal{R}_{\text{P}}^{\text{AH}} = \mathcal{C}_{\text{P}}^{\text{AH}}/(4 \pi M_{T}), \\
    & \mathcal{R}_{\text{E}}^{\text{AH}} = \mathcal{C}_{\text{E}}^{\text{AH}}/(4 \pi M_{T}),
\end{align}
defined with respect to the total mass and the following ones with respect to the Hawking mass:
\begin{align}
    & \widetilde{\mathcal{R}}_{\text{P}}^{\text{AH}} = \mathcal{C}_{\text{P}}^{\text{AH}}/(4 \pi M_{\mathcal{H}}), \\
    & \widetilde{\mathcal{R}}_{\text{E}}^{\text{AH}} = \mathcal{C}_{\text{E}}^{\text{AH}}/(4 \pi M_{\mathcal{H}}),
\end{align}
which serve us to test the hoop conjecture. In general, the latter will be greater than the former since $ M_{\mathcal{H}} \leq M_{T}$, with the equality holding for nonrotating configurations. According to the hoop conjecture, these ratios should be smaller than one, and we will explore which of the two masses should be considered in this situation. Ref.~\cite{Senovilla2007} touches on this topic arguing that it is to be expected that we should use a quantity that 
takes rotation into account. 

This is motivated by considering the Kerr geometry. For the Kerr black hole, we can compute the equatorial hoop:
\begin{align}
    & \mathcal{C}_{\text{E}} = 4 \pi M_{T} = 4 \pi M_{\mathcal{H}} \sqrt{1 + \frac{J^2_{\mathcal{H}}}{4 M_{\mathcal{H}}^4}},
\end{align}
where we have expressed it in terms of both the total mass (which for Kerr agrees with the ADM mass) and the irreducible mass. 
From this, we clearly see that the conjecture would be violated if we interpreted $M_{\mathcal{H}}$ as the candidate to enter the hoop inequality for the Kerr black hole. 
Specifically, for extremal Kerr with $J_\mathcal{H} = M_{T}^2$ the extremal equatorial hoop is
\begin{align}
    \mathcal{C}_{\text{E}}\,\big|_{ex} =4 \sqrt{2} \,\pi M_{\mathcal{H}}, \label{Eq:KerrExtremalHoop}
\end{align}
which violates the \emph{only if} part of the hoop conjecture by $\sim 41 \%$ if using $M_{\mathcal{H}}$.
This means that $M_T$ is, at least for the Kerr black hole, a more sensible candidate. Below we delve into this by comparing both of them for other configurations that are not stationary and present a more complicated structure.

Motivated by its validity for Kerr--Newman--(anti-)de~Sitter black holes, Hod~\cite{Hod2020} has conjectured that every black hole satisfies
\begin{align}
    \mathcal{C}_s \geq 4 \pi M_{\text{H}},
\end{align}
where $\mathcal{C}_{s}$ is the circumference of the smallest ring that can engulf the horizon in every direction. Hence, in our case it means that 
\begin{align}
    \text{max} (\widetilde{\mathcal{R}}_{\text{P}}^{\text{AH}}, \widetilde{\mathcal{R}}_{\text{E}}^{\text{AH}}) \geq 1. 
\end{align}
We check this conjecture for all of our rotating solutions.

The second relevant surface is the one which is outside the matter content, i.e., it does not intersect it, and minimizes the length of the curves in Eqs.~\eqref{Eq:EquatorialCircumf}--\eqref{Eq:PolarCircumf}. This will be either a geodesic or the curve that sits exactly on top of the boundary of the region enclosing the matter content. We call those equatorial and polar circumferences $\mathcal{C}^{\text{min}}_E$ and $\mathcal{C}^{\text{min}}_P$, respectively. 

For the equatorial case, we can consider Eq.~\eqref{Eq:EquatorialCircumf}, which depends only on the undetermined quantity $R(\pi/2)$, i.e., the radial distance of the surface from the origin in the $z = 0$ plane (which is constant along the azimuthal direction due to axisymmetry). The problem therefore reduces to finding the value of $R(\pi/2)$ that optimises Eq.~\eqref{Eq:EquatorialCircumf}. We do this numerically by evaluating the equation for different values of $R(\pi/2)$ and identifying the optimal one.

For the polar case, additional computation is needed. In fact, we can directly apply the Euler-Lagrange equations to derive the expression for the surface $R(\theta)$ that makes extremal the functional~\eqref{Eq:PolarCircumf}. The resulting equation is
\begin{align}
    R^{\prime \prime}  - 2 \frac{\left(R'\right)^2}{R} - R + 2 \left[ 1 + \left( \frac{R'}{R} \right)^2 \right] \left[ \frac{\partial_\theta \psi}{\psi} R' - \frac{\partial_R \psi}{\psi} R^2 \right] = 0,
    \label{Eq:PolarGeodesics}
\end{align}
where the conformal factor is evaluated on the surface. This equation also corresponds to the equation for a 3D geodesic curve in a non-affine parametrization, using the coordinate $\theta$ as the parameter, as those are the curves that (locally) minimize the length between two points. In general, smoothness of the surface (and hence the curve) requires that the function $R(\theta)$ is even and hence $R' (0) = 0$. Also, again reflection symmetry enforces that the derivative also vanishes at $R' (\pi/2) = 0$. The structure of the equation and the boundary conditions are essentially the same as the one for the horizon, hence it can also be solved by shooting or relaxation. In this case, we have solved it only with a relaxation algorithm applying the same scheme as we have done for the horizon. 

If no solution exists satisfying this boundary conditions, or the surface $R(\theta)$ intersects the matter content, it means that the polar curve outside the matter content that minimizes the length is the one that sits on the surface of the matter content and we consider it. We find again convenient to introduce the following ratios:
\begin{align}
    & \mathcal{R}_{\text{P}}^{\text{min}} = \mathcal{C}_{\text{P}}^{\text{min}}/(4 \pi M_{\text{cont}}), \\
    & \mathcal{R}_{\text{E}}^{\text{min}} = \mathcal{C}_{\text{E}}^{\text{min}}/(4 \pi M_{\text{cont}}),
\end{align}
where we have introduced the mass contained inside the surface as the same integral defined in Eq.~\eqref{Eq:AsymptM} where the second term is integrated only inside the matter content, i.e., the rotational part of the integrand that extends up to infinity is only integrated inside the surface
\begin{align}
    M_{\text{cont}} = \, \, \,  2 \pi &  \int_{M} \dd r \, \dd z \, r \rho_{\text{N}} (r,z) \,  \nonumber \\
     +  \frac{1}{4} &  \int_{M} \dd r \, \dd z \, r \psi^{-7} \left[   \left( \partial_r  \mathcal{W} \right)^2 + \frac{ \mathcal{W}^2}{r^2} - 2 \frac{ \mathcal{W} \partial_r  \mathcal{W}}{r} +  (\partial_z  \mathcal{W})^2 \right].
    \label{Eq:ContM}
\end{align}
According to the hoop conjecture, ${\rm max}\left(\mathcal{R}_{\text{P}}^{\text{min}},\mathcal{R}_{\text{E}}^{\text{min}}\right)\geq 1$, as otherwise a horizon would be expected. We notice that in the nonrotating case the ADM mass reduces to the integral over the 3D region with a nonvanishing density, and hence it can be understood also as a quasilocal quantity and can be used for the hoop conjecture tests, as it is done in Ref.~\cite{Nakamura1988}. However, in the rotating case, the ADM mass is an integral across the full 3D spatial hypersurface, and it satisfies $M_{\text{ADM}} > M_{\text{cont}}$. Hence, it leads to lower values of the ratios that are closer to violate the hoop conjecture.
\subsection{Naked singularities and the hoop conjecture}
\label{Subsec:OverSpinning}
The toy model that originally motivated the hoop conjecture, an infinitely long cylinder of matter undergoing gravitational collapse, forms a naked singularity~\cite{Thorne1972}, although it is clear that such a configuration is physically unrealistic. This leaves open the question of what happens with very massive configurations that, according to the hoop conjecture, are not compacted enough in some directions to form a horizon. In such situations, if a singularity forms, it would be a naked singularity, i.e., not covered by a horizon, which would constitute a violation of the weak cosmic censorship conjecture~\cite{Penrose1973,Wald1997}. This suggests that there is a kind of trade-off between the hoop conjecture and the weak cosmic censorship conjecture, in cases where no horizon forms. 

One of the main analyses that inspired the present work, the one by Nakamura, Shapiro, and Teukolsky~\cite{Nakamura1988}, considers momentarily static ellipsoids of dust with variable eccentricities and solves the Hamiltonian constraint to find initial data. Furthermore, they look for apparent horizons in those initial data. They show that above a given eccentricity, the configurations are horizonless and hence satisfy the hoop conjecture, since one cannot pass a sufficiently small hoop around the matter in every direction. Furthermore, they show that in the infinite eccentricity limit, the configurations become singular, with such singularity being naked. 

These results suggested that a naked singularity might form dynamically if collapse drives the configuration towards increasingly large eccentricity. Shapiro and Teukolsky~\cite{Shapiro1991} subsequently evolved highly prolate spheroids of collisionless matter and found no apparent horizon, while curvature invariants appeared to grow without bound, which they interpreted as evidence for the formation of a naked singularity. A later study~\cite{Yoo2016} also found singular behaviour without apparent horizons in similar configurations. The conclusion that the singularity is naked, however, was questioned on the grounds that apparent horizons are slicing dependent: even Schwarzschild admits slicings with no trapped surfaces~\cite{Wald1991}. Indeed, East~\cite{East2019} revisited the problem with different slicing and coordinate conditions and found that a black hole does form and eventually swallows all the matter, so that no naked singularity appears. In his simulations, the matter is compacted in all directions before the horizon forms, consistently with the hoop conjecture, and the collapse is accompanied by strong gravitational radiation.

A different criticism was raised by Rendall~\cite{Rendall1992}, who argued that the particle methods used in Ref.~\cite{Shapiro1991}, and by extension in Refs.~\cite{Yoo2016,East2019}, effectively describe dust, which is prone to singular behaviour, rather than the better-behaved Einstein--Vlasov system. Dust is recovered from the Vlasov equation only in a singular, distributional limit, in which the phase-space distribution is a delta function in momentum. More recently, Ref.~\cite{Ames2023} studied the same problem for the Einstein--Vlasov system and reached conclusions similar to East's: an apparent horizon enclosing all the matter always forms, so weak cosmic censorship is not violated, and the hoop conjecture is satisfied within $12\%$, in agreement with Ref.~\cite{Nakamura1988}.

On top of this, rotation adds an extra ingredient into the picture and we only give a first step here towards understanding its impact. One expects that, for configurations that collapse to a black hole, the angular momentum and the ADM masses should satisfy the relation \mbox{$\chi_{\text{ADM}} \leq 1 $}. This ensures that the end state of the collapse cannot be a super extremal Kerr black hole, which contains a naked singularity, and would therefore violate the weak cosmic censorship conjecture~\cite{Penrose1973,Wald1997}. For instance, this is proved in vacuum in Ref.~\cite{Dain2006}\footnote{
For ordinary bodies, a related inequality bounds the angular momentum by the size, $J\lesssim\mathcal{R}^2$, where $\mathcal{R}$ is a suitable measure of the size. Plausibility arguments for nonvacuum spacetimes, together with a proof for constant-density bodies, were given in Ref.~\cite{Dain2013}. For bodies close to forming a black hole, $\mathcal{R} \sim M$, and the inequality suggests $\chi_{\text{ADM}} \lesssim \mathcal{O}(1)$.}. Here we also explore whether it is possible to find reasonable initial data (that do not violate energy conditions) that violate this inequality. 

A related inequality is the Penrose inequality~\cite{Penrose1973}, which states that the ADM mass is bounded below by the irreducible mass of the outermost apparent horizon, \mbox{$M_{\text{ADM}} \geq M_{\mathcal{H}}$}. Penrose's heuristic argument is that, if the spacetime settles down to a Kerr black hole and cosmic censorship holds, the area theorem together with the decrease of the Bondi mass implies this bound, so a violation would signal a naked singularity. The inequality has been proved in the Riemannian (time-symmetric) case, first for a connected horizon~\cite{Huisken2001} and then for horizons with several components~\cite{Bray2001}; the general case remains open. These theorems apply to our data as well: on a maximal slice (i.e. $K=0$), the Hamiltonian constraint gives ${}^{3}R = 16\pi\rho + K_{ij}K^{ij} \geq 0$, and our apparent horizons are minimal surfaces. Accordingly, we find $M_{\text{ADM}} \geq M_{\mathcal{H}}$ in all cases, which provides a consistency check of our numerics. Including angular momentum, one may ask whether $M_{\text{ADM}} \geq M_T$, with $M_T$ the Christodoulou mass of Eq.~\eqref{Eq:TotalMass}. This quantity has been proposed as a quasilocal mass for dynamical axisymmetric black holes~\cite{Dain:2017jkj}, and Penrose-like inequalities with angular momentum have been studied in Ref.~\cite{Anglada:2017ryp}. All our horizon configurations satisfy it. 

\section{Numerical Results}
\label{Sec:Numeric}
We consider three classes of configurations. First, nonrotating spheroids similar to those of Ref.~\cite{Nakamura1988}, except that we use smooth density profiles instead of discontinuous ones, which is more convenient numerically. Second, configurations with a constant rescaled angular velocity $v_{\text{N}}^{\phi}$, which we refer to as rigidly rotating. Third, differentially rotating configurations, in which $v_{\text{N}}^{\phi}$ varies in space.

\subsection{Nonrotating configurations}
\label{Subsec:non-rotatingConfigs}
\begin{figure}[htbp]
    \centering
    \includegraphics[width=0.8\textwidth]{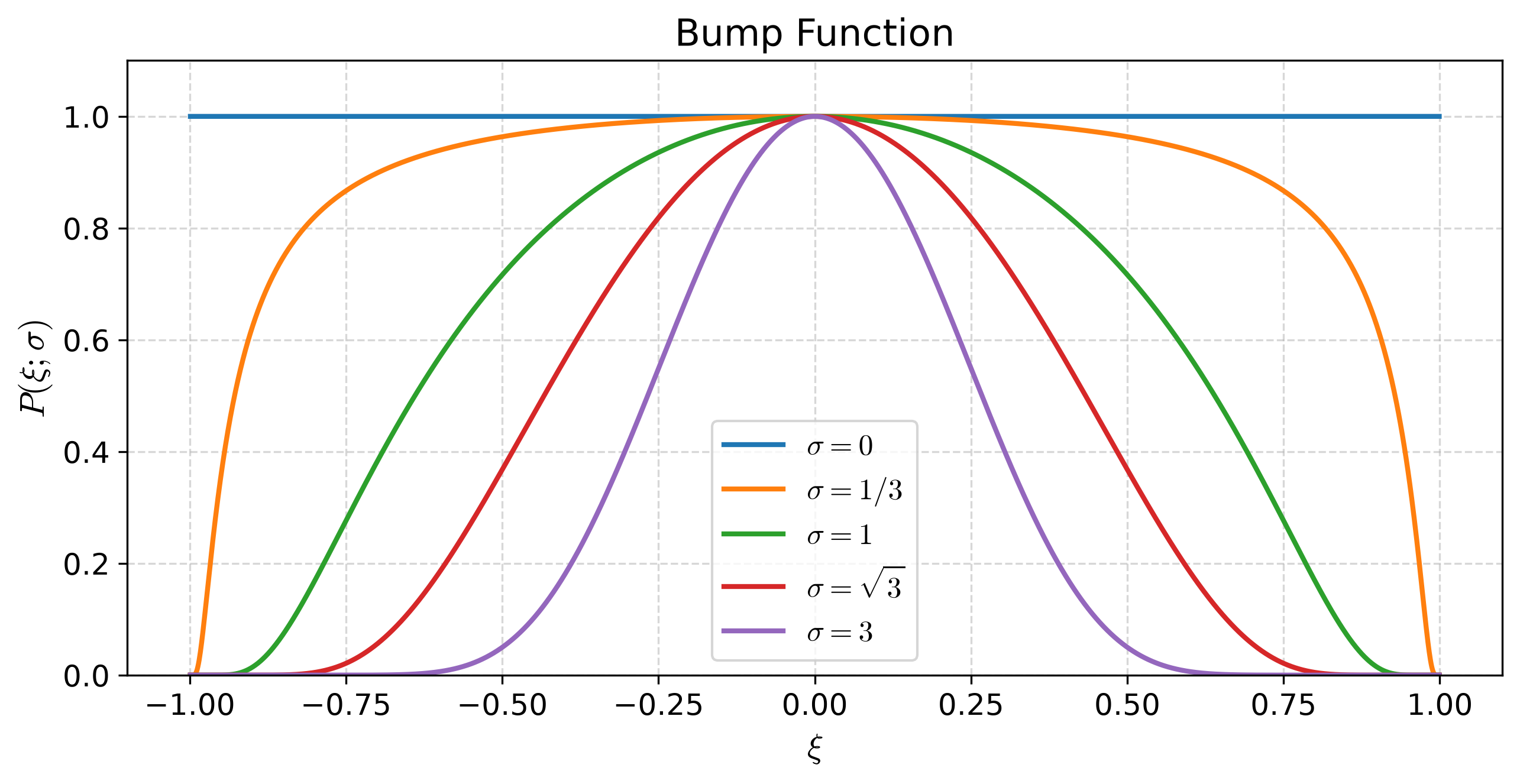}
    \caption{The bump function $P(\xi;\sigma) = e^{\sigma^2 \xi^2/(\xi^2-1)}$ for varying $\sigma$. Note that the red curve is our default choice.}
    \label{fig:bump_profile}
\end{figure}

We consider nonrotating profiles in this section, hence we take $v^{\phi} = 0$ which implies also that $\mathcal{W} = 0$. We conveniently introduce the dimensionless coordinate $\xi$
\begin{align}
    \xi^2 = \frac{z^2}{a^2} + \frac{r^2}{b^2},
\end{align}
such that the interior region of an ellipsoid with semi-axis $a$ and $b$ corresponds to the region $\xi^2 < 1$. We generically define the eccentricity of these objects as:
\begin{align}
    e := \sqrt{1 - \left(\frac{\text{min}(a,b)}{\text{max} (a,b)} \right)^2}.
\end{align}
In terms of the $\xi$ coordinate, we can express the density profile as
\begin{align}
    \rho_{\text{N}} = \rho_0 P(\xi; \sigma),
    \label{Eq:BumpDensity}
\end{align}
with $\rho_0$ a parameter representing the maximum value of the density, and the function $P(\xi;\sigma)$ is a $\mathcal{C}^{\infty}$ bump function whose value is one at the centre of the ellipse and tends to zero at the border, smoothly interpolating with a zero value outside of the ellipse. Its explicit expression is
\begin{align}
    P(\xi;\sigma) = 
    \begin{cases}
        \mathrm{exp}\left({\frac{\sigma^2 \xi^2}{\xi^2-1}}\right), & \xi < 1,\\
        0, & \xi \geq 1,
    \end{cases}  
    \label{eq:bump}
\end{align}
where $\sigma$ is a parameter that controls how fast the density profile decays at the border of the ellipse; the profile is shown in Fig.~\ref{fig:bump_profile}. Unless otherwise stated we take $\sigma = \sqrt{3}$.

We also recall that we are fixing the volume of the ellipsoids to $V = 4 \pi/3$ in order to remove the overall scale as a degree of freedom. Consequently, the parameters $a$ and $b$ must satisfy the constraint $a b^2 = 1$. 

This profile corresponds to having all the matter content inside a ellipsoidal shape which can be either oblate (pancake-like) in the case in which $a<b$ or prolate (baguette-like), in the case in which $a>b$. The limit $a=b$ corresponds to a spherical distribution. In the limit in which the parameter $\sigma \to 0$, it approaches a discontinuous characteristic function with support only inside the ellipsoid, which is actually the one considered by Nakamura, Shapiro and Teukolsky~\cite{Nakamura1988}. Given that we are performing a numerical integration, we find it convenient to work with smooth density profiles. We find a qualitative agreement with their conclusions, as the configurations that we take are similar to theirs.

We report here five different runs in which we keep all the parameters fixed but one. Namely, we have considered:
\begin{enumerate}

    \item Spherically symmetric solutions in which we sweep over the only free parameter, the density $\rho_0$.

    \item Oblate ellipsoids of fixed eccentricity $e$ in which we sweep over the density parameter $\rho_0$.

    \item Prolate ellipsoids of fixed eccentricity $e$ in which we sweep over the density parameter $\rho_0$. 

    \item Oblate ellipsoids of fixed density $\rho_0$ in which we sweep over the eccentricity $e$. 

    \item Prolate ellipsoids of fixed density $\rho_0$ in which we sweep over the eccentricity $e$.
\end{enumerate}
We have performed other simulations but here we report the values that are close to the transition from having to not having an apparent horizon, as those critical cases are the ones that probe the hoop conjecture in the most extremal situations. For fixed density, increasing the eccentricity would make the apparent horizon disappear at some point, whereas for fixed eccentricity increasing the density would make an apparent horizon appear.

\begin{table}[htbp] 
    \centering
    \setlength{\tabcolsep}{4pt}
    \begin{tabular}{cccccccc}
        \hline\hline
        $\rho_0$  & AH? & $M_{\mathcal{H}}$ & $M_{\text{ADM}}$ & $\mathcal{R}_{\text{P}}^{\text{min}}$ & $\mathcal{R}_{\text{E}}^{\text{min}}$ &  $\mathcal{R}_{\text{P}}^{\text{AH}}$ & $\mathcal{R}_{\text{E}}^{\text{AH}}$ \\
        \hline \hline
        3.00 & $\cross$ & - & 1.47553 & 1.02321  & 1.02321 & - & - \\
        \hline
        3.02 & $\cross$ & - & 1.48537 & 1.02219 & 1.02219 & - & - \\
        \hline
        3.04 & $\cross$ & - & 1.49520 & 1.02121 & 1.02121 & - & - \\
        \hline
        3.06 & $\checkmark$ & 1.50407 & 1.50504 & 1.02025 & 1.02025 & 0.999987 & 0.999987 \\
        \hline
        3.08 & $\checkmark$ & 1.51409 & 1.51488 & 1.01932 & 1.01932 & 0.999987 & 0.999987 \\
        \hline
        3.10 & $\checkmark$ & 1.52406 & 1.52472 & 1.01842 & 1.01842 & 0.999987 & 0.999987 \\
        \hline
        \hline
    \end{tabular}
    \caption{ Spherical configurations with varying density parameter $\rho_0$. ``AH?'' indicates whether an apparent horizon is present; $M_{\mathcal{H}}$ is its irreducible (Hawking) mass and $M_{\text{ADM}}$ the ADM mass. $\mathcal{R}^{\text{AH}}_{\text{P,E}}$ are the polar and equatorial circumferences of the horizon divided by $4\pi M_{\mathcal{H}}$, and $\mathcal{R}^{\text{min}}_{\text{P,E}}$ are the minimal polar and equatorial circumferences enclosing the matter divided by $4\pi M_{\text{cont}}$ (see Sec.~\ref{Subsec:AxialSym}). Without rotation $M_T = M_{\mathcal{H}}$, so the tilded ratios are not needed. In spherical symmetry the polar and equatorial circumferences coincide, and moreover $\mathcal{R}^{\text{AH}} = 1$ identically, since the horizon circumference equals $2\pi$ times its areal radius, which equals $4\pi M_{\mathcal{H}}$. The deviation of $\mathcal{R}^{\text{AH}}$ from unity, about $10^{-5}$, therefore measures our numerical accuracy. As $\rho_0$ increases, $M_{\mathcal{H}}$ approaches $M_{\text{ADM}}$ because the horizon encloses a growing fraction of the matter.}
    \label{Tab:SpheresDensities}
\end{table}
\begin{figure}[htbp]
    \centering
    \includegraphics[width=0.6\textwidth]{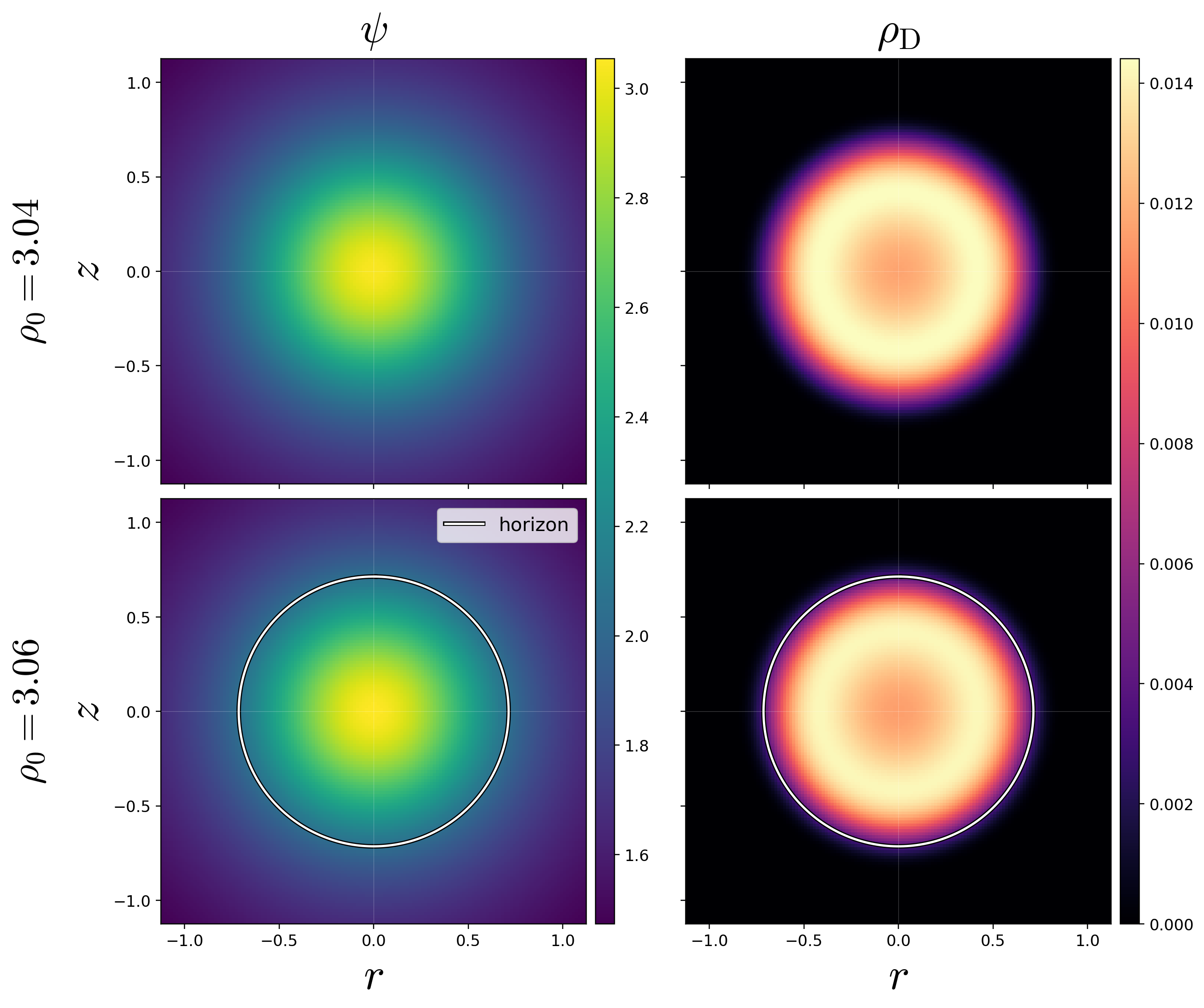}
    \caption{Here we compare two spherical configurations from the Table~\ref{Tab:SpheresDensities}, namely \mbox{$\rho_0 = 3.04$} in the upper two panels (no horizon) and $\rho_0 = 3.06$ in the lower two panels. For both, the conformal factor is shown on the left and the physical source $\rho_{\text{D}}$ on the right.}
    \label{Fig:SphereNHvsAH}
\end{figure}

The results for the first case are reported in Table~\ref{Tab:SpheresDensities}, where we run simulations for the spheres varying the $\rho_0$ parameter in the interval $(3.00-3.10)$ in steps of $0.02$. The horizon appears for $\rho_0 \geq 3.06$. We note that in this case the exterior metric is simply Schwarzschild. 

\begin{table}[htbp] 
    \centering
    \setlength{\tabcolsep}{4pt}
    \begin{tabular}{cccccccc}
        \hline\hline
        $\rho_0$  & AH? & $M_{\mathcal{H}}$ & $M_{\text{ADM}}$ & $\mathcal{R}_{\text{P}}^{\text{min}}$ & $\mathcal{R}_{\text{E}}^{\text{min}}$ &  $\mathcal{R}_{\text{P}}^{\text{AH}}$ & $\mathcal{R}_{\text{E}}^{\text{AH}}$ \\
        \hline 
        \hline
        5.0 & $\cross$ & - & 2.45902 & 1.43543  & 1.08465 & - & - \\
        \hline
        5.1 & $\checkmark$ & 2.50342 & 2.5082 & 1.44046  & 1.08046 & 0.958659 & 1.13537 \\
        \hline
        5.2 & $\checkmark$ & 2.55366 & 2.55738 & 1.44567  & 1.07656 & 0.961572 & 1.12143 \\
        \hline
        5.3 & $\checkmark$ & 2.60355 & 2.60656 & 1.45107  & 1.07295 & 0.963656 & 1.11155 \\
        \hline
        5.4 & $\checkmark$ & 2.65323 & 2.65574 & 1.45664  & 1.06959 & 0.965348 & 1.10368 \\
        \hline
        5.5 & $\checkmark$ & 2.70276 & 2.70492 & 1.46237 & 1.06649 & 0.966798 & 1.09708 \\
        \hline
        \hline
    \end{tabular}
    \caption{Results for oblate ellipsoids of fixed eccentricity $ e = 0.99$ with varying densities. The labels of the columns are the same as the ones in Table~\ref{Tab:SpheresDensities}.}
    \label{Tab:OblateDensities}
\end{table}
\begin{figure}[htbp]
    \centering
    \includegraphics[width=0.6\textwidth]{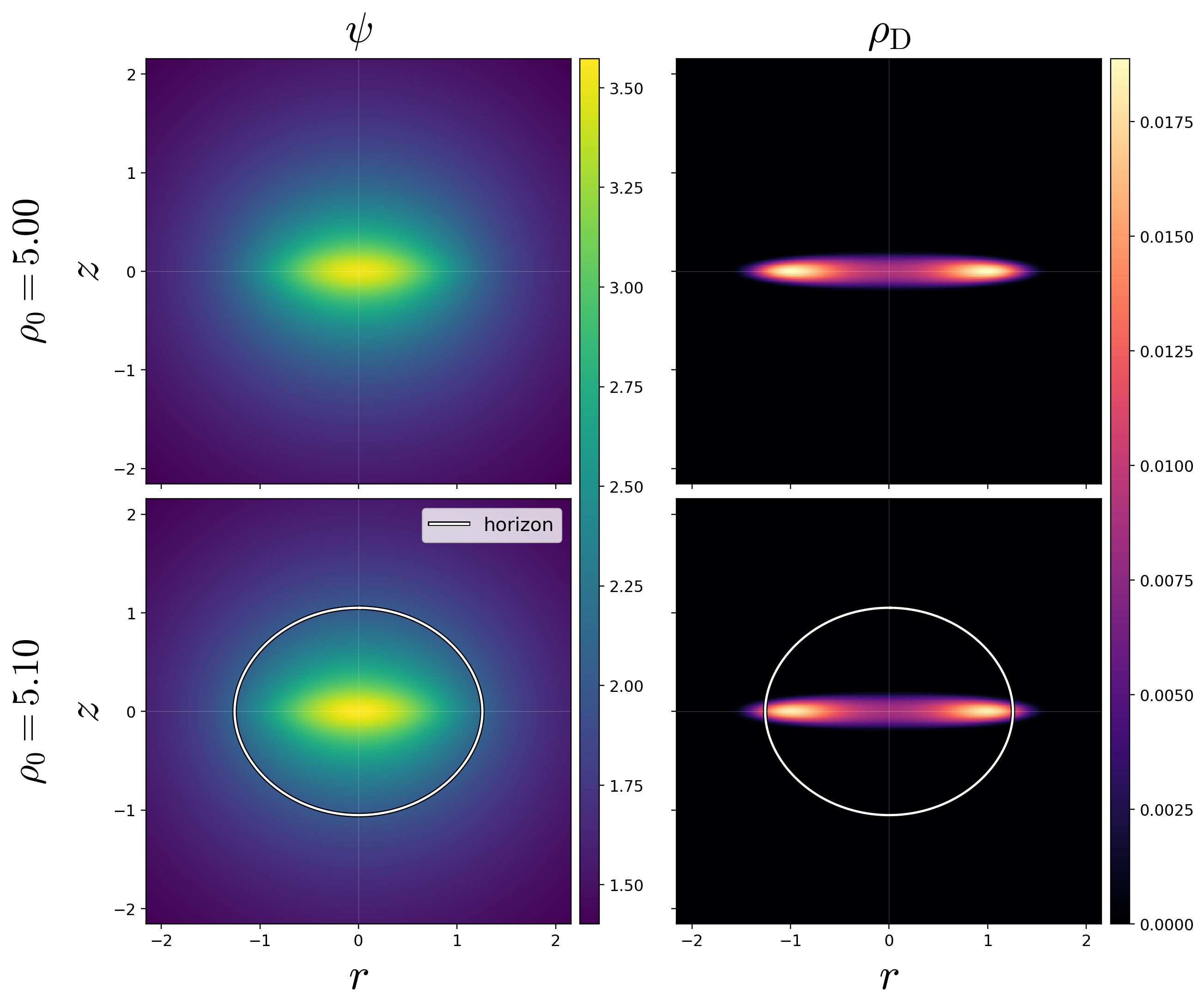}
    \caption{Here we compare two oblate configurations from the Table~\ref{Tab:OblateDensities}, namely $\rho_0 = 5.0$ in the upper two panels (no horizon) and $\rho_0 = 5.1$ in the lower two panels. For both, the conformal factor is shown on the left and the physical source $\rho_{\text{D}}$ on the right.}
    \label{Fig:PancakeNHvsAH}
\end{figure}

The results for the second case are reported in Table~\ref{Tab:OblateDensities}, where we run simulations for oblate ellipsoids of fixed eccentricity $e = 0.99$ varying the $\rho_0$ parameter in the interval $(5.0-5.5)$ in steps of $0.1$. The horizon appears for $\rho_0 \geq 5.1$.

\begin{table}[htbp] 
    \centering
    \setlength{\tabcolsep}{4pt}
    \begin{tabular}{cccccccc}
        \hline\hline
        $\rho_0$  & AH? & $M_{\mathcal{H}}$ & $M_{\text{ADM}}$ & $\mathcal{R}_{\text{P}}^{\text{min}}$ & $\mathcal{R}_{\text{E}}^{\text{min}}$ &  $\mathcal{R}_{\text{P}}^{\text{AH}}$ & $\mathcal{R}_{\text{E}}^{\text{AH}}$ \\
        \hline \hline
        6.0 & $\cross$ & - & 2.95194 & 1.81007 & 0.665319 & - & - \\
        \hline
        6.2 & $\cross$ & - & 3.05034 & 1.81485 & 0.671444 & - & - \\
        \hline
        6.4 & $\cross$ & - & 3.14874 & 1.82046 & 0.677747 & - & - \\
        \hline
        6.6 & $\cross$ & - & 3.24714 & 1.82681 & 0.684211 & - & - \\
        \hline
        6.8 & $\cross$ & - & 3.34554 & 1.83385 & 0.690823 & - & - \\
        \hline
        7.0 & $\checkmark$ & 3.41572 & 3.44393 & 1.84151 & 0.69757 & 1.13129 & 0.851426 \\
        \hline\hline
    \end{tabular}
    \caption{Results for prolate ellipsoids of fixed eccentricity $ e = 0.99$ with varying densities. The labels of the columns are the same as the ones in Table~\ref{Tab:SpheresDensities}.}
    \label{Tab:ProlateDensities}
\end{table}
\begin{figure}[htbp]
    \centering
    \includegraphics[width=0.6\textwidth]{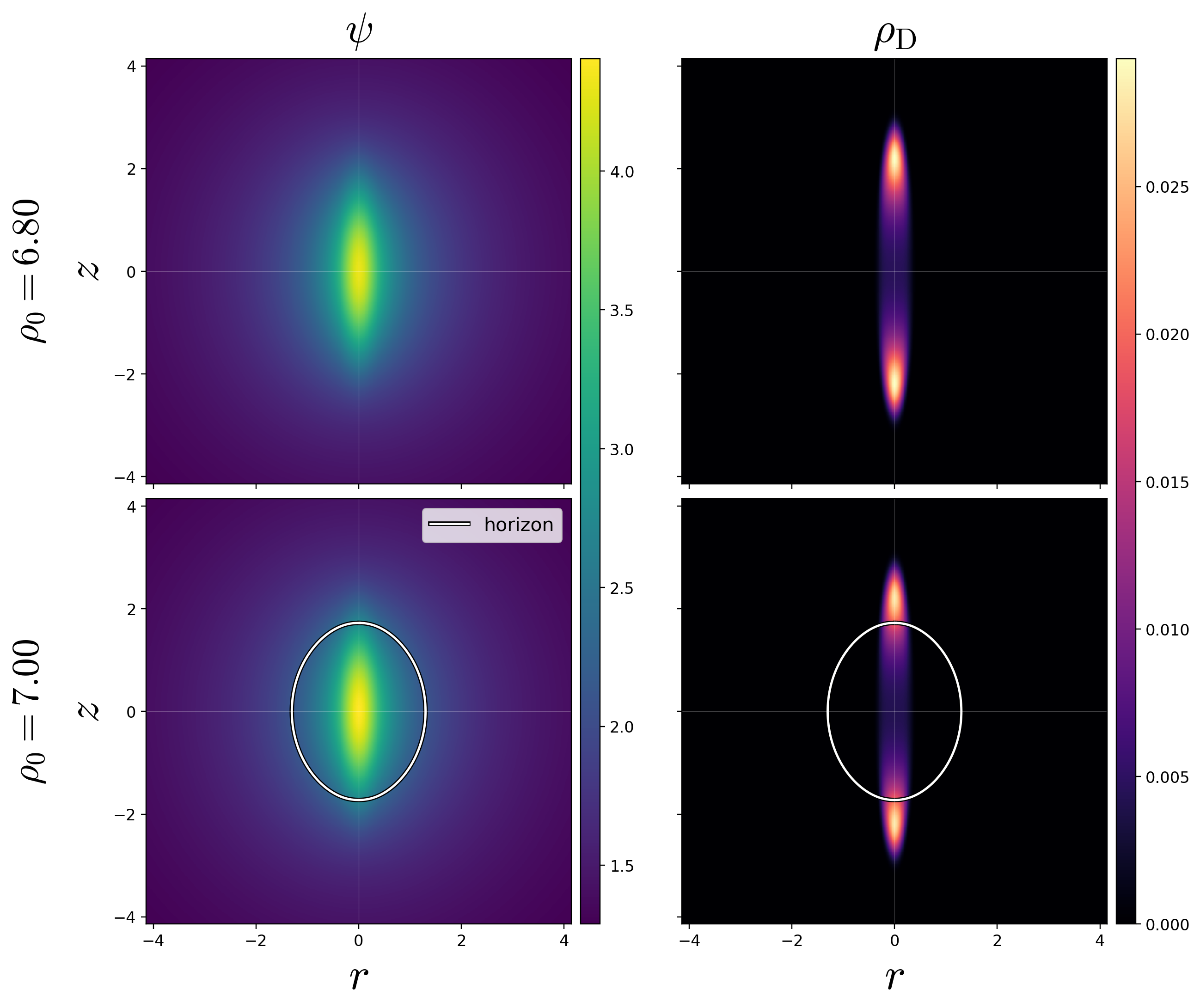}
    \caption{Here we compare two prolate configurations from the Table~\ref{Tab:ProlateDensities}, namely $\rho_0 = 6.8$ in the upper two panels (no horizon) and $\rho_0 = 7.0$ in the lower two panels. For both, the conformal factor is shown on the left and the physical source $\rho_{\text{D}}$ on the right.}
    \label{Fig:BaguetteNHvsAH}
\end{figure}

The results for the third case are reported in Table~\ref{Tab:ProlateDensities}, where we run simulations for prolate ellipsoids of fixed eccentricity $e = 0.99$ varying the $\rho_0$ parameter in the interval $(6.0-7.0)$ in steps of $0.2$. The horizon appears for $\rho_0 \geq 7.0$. 

\begin{table}[htbp] 
    \centering
    \setlength{\tabcolsep}{4pt}
    \begin{tabular}{cccccccc}
        \hline\hline
        $e$  & AH? & $M_{\mathcal{H}}$ & $M_{\text{ADM}}$ & $\mathcal{R}_{\text{P}}^{\text{min}}$ & $\mathcal{R}_{\text{E}}^{\text{min}}$ &  $\mathcal{R}_{\text{P}}^{\text{AH}}$ & $\mathcal{R}_{\text{E}}^{\text{AH}}$ \\
        \hline \hline
        0.93 & $\checkmark$ & 1.96585 & 1.96718 & 1.14596 & 1.0615  & 0.97037 & 1.08011 \\
        \hline
        0.94 & $\checkmark$ & 1.96537 & 1.96718 & 1.16609 & 1.06655  & 0.967854 & 1.08966 \\
        \hline
        0.95 & $\checkmark$ & 1.96443 & 1.96718 & 1.19051 & 1.07286  & 0.964155 & 1.10429 \\
        \hline
        0.96 & $\cross$ & - & 1.96718 & 1.22093 & 1.08109  & - & - \\
        \hline
        0.97 & $\cross$ & - & 1.96718 & 1.26027 & 1.09263  & - & - \\
        \hline
        0.98 & $\cross$ & - & 1.96717 & 1.31441 & 1.1108  & - & - \\
        \hline
        0.99 & $\cross$ & - & 1.96717 & 1.39902 & 1.14763  & - & - \\
        \hline \hline
    \end{tabular}
    \caption{Results for oblate ellipsoids of fixed density $\rho_0 = 4.0$ and varying eccentricity $e$. The labels of the columns are the same as the ones in Table~\ref{Tab:SpheresDensities}.}
    \label{Tab:OblateEccentricities}
\end{table}

The results for the fourth case are reported in Table~\ref{Tab:OblateEccentricities}, where we run simulations for oblate ellipsoids of fixed density parameter $\rho_0  = 4.0$ varying the eccentricity in the interval $(0.93-0.99)$ in steps of $0.01$. The horizon is present for $e \leq 0.95$.

\begin{table}[htbp] 
    \centering
    \setlength{\tabcolsep}{4pt}
    \begin{tabular}{cccccccc}
        \hline\hline
        $e$  & AH? & $M_{\mathcal{H}}$ & $M_{\text{ADM}}$ & $\mathcal{R}_{\text{P}}^{\text{min}}$ & $\mathcal{R}_{\text{E}}^{\text{min}}$ &  $\mathcal{R}_{\text{P}}^{\text{AH}}$ & $\mathcal{R}_{\text{E}}^{\text{AH}}$ \\
        \hline \hline
        0.93 & $\checkmark$ & 2.45702 & 2.45908 & 1.25943 & 0.895402 & 1.04531 & 0.931973 \\
        \hline
        0.94 & $\checkmark$ & 2.4562 & 2.45908 & 1.29351 & 0.879791 & 1.05236 & 0.923937 \\
        \hline
        0.95 & $\checkmark$ & 2.45465 & 2.45909 & 1.3362 & 0.859942 & 1.0622 & 0.913199 \\
        \hline
        0.96 & $\checkmark$ & 2.45120 & 2.45910 & 1.3917 & 0.833721 & 1.07745 & 0.897203 \\
        \hline
        0.97 & $\cross$ & - & 2.45912 & 1.46813 & 0.797095 & - & - \\
        \hline
        0.98 & $\cross$ & - & 2.45914 & 1.58419 & 0.741084 & - & - \\
        \hline
        0.99 & $\cross$ & - & 2.45920 & 1.80257 & 0.638302 & - & - \\
        \hline\hline
    \end{tabular}
    \caption{Results for prolate ellipsoids of fixed density $\rho_0 = 5.0$ and varying eccentricity $e$. The labels of the columns are the same as the ones in Table~\ref{Tab:SpheresDensities}.}
    \label{Tab:ProlateEccentricities}
\end{table}

The results for the fifth case are reported in Table~\ref{Tab:ProlateEccentricities}, where we run simulations for prolate ellipsoids of fixed density parameter $\rho_0  = 5.0$ varying the eccentricity in the interval $(0.93-0.99)$ in steps of $0.01$. The horizon is present for $e \leq 0.96 $.

We find a good qualitative agreement with the results by Nakamura, Shapiro and Teukolsky~\cite{Nakamura1988}. We find that for all the configurations with an apparent horizon, the \emph{only if} part of the inequality entering the conjecture is satisfied in all the cases within $14 \%$. We also corroborate that no universal numerical coefficient can exist for the \emph{if} part of the conjecture, as for the same values of the minimum ratio some cases have a horizon whereas others do not. Remarkably, Hod's inverse hoop conjecture~\cite{Hod2020} is satisfied in all the configurations we have studied. In the spherically symmetric case the bound is saturated, as it must be, since $\mathcal{C}^2/\pi = \mathcal{A}$ identically, and our numerical results reproduce this equality within numerical precision. 

\subsection{Rigidly rotating configurations}
\label{Subsec:RigidRotation}
For the rigidly rotating configurations we keep the same density profile specified in Eq.~\eqref{Eq:BumpDensity}, but now include a nonvanishing profile for the angular velocity $v^\phi$ which is:
\begin{align}
    v^{\phi}_{\text{N}} = v_0 = \text{const},
\end{align}
which we refer to as rigid rotation, since the rescaled angular velocity is uniform. The physical angular velocity, $v^{\phi} = v_0\, \psi^{-5}$, is not uniform, as discussed below. Since only the product $\rho_{\text{N}}\, v_{\text{N}}^{\phi}$ enters the right-hand side of Eq.~\eqref{Eq:WPRescequation}, the source remains compactly supported for any bounded $v_{\text{N}}^{\phi}$, so no additional scale needs to be introduced in the velocity profile.

From Eq.~\eqref{Eq:vNdef}, the physical linear velocity of the fluid is $v^{\phi}\psi^2 r = v_0\, r\, \psi^{-3}$. Since $\psi$ decreases outwards, the fluid moves faster near the surface of the object than near its centre, both because of the factor $r$ and because of the decreasing conformal factor.

We recall, as discussed in Sec.~\ref{Subsec:3+1}, that the solutions should be such that the product $v^\phi \psi^2 r < 1$ everywhere, as otherwise we would have an effective superluminality. We have made sure that this bound is satisfied everywhere for each of our solutions. 

We have run three series of simulations in this case. The first results with rigid rotation are reported in Table~\ref{Tab:SpheresRigidRotation}, where we run simulations for the spheres varying the $v_0$ parameter in the interval $(0-4)$ in steps of $1$; the density is fixed to $\rho_0 = 3.1$. The horizon disappears for $v_0 \geq 4$. Thus, at fixed $\rho_0$, increasing the rotation removes the apparent horizon, even though both the ADM mass and the Christodoulou mass increase. Note, however, that at fixed rescaled density the physical density also changes with $v_0$, so this sequence does not isolate the effect of rotation at fixed physical matter content.
\begin{table}[htbp] 
    \centering
    \setlength{\tabcolsep}{3.5pt}
    \renewcommand{\arraystretch}{1.2}
    \begin{tabular}{cccccccccccc}
        \hline\hline
        $v_0$  & AH? & $M_{\text{ADM}}$ & $\chi_{\text{ADM}}$ & $M_{T}$ & $\chi_{\mathcal{H}}$ & $\mathcal{R}_{\text{P}}^{\text{min}}$ & $\mathcal{R}_{\text{E}}^{\text{min}}$ &  $\mathcal{R}_{\text{P}}^{\text{AH}}$ & $\mathcal{R}_{\text{E}}^{\text{AH}}$ &   $\widetilde{\mathcal{R}}_{\text{P}}^{\text{AH}}$ & $\widetilde{\mathcal{R}}_{\text{E}}^{\text{AH}}$  \\
        \hline \hline
        0.0 & $\checkmark$ & 1.5247 & 0.0000 & 1.5241 & 0.0000 & 1.0184 & 1.0184 & 1.0000 & 1.0000 & 1.0000 & 1.0000 \\
        1.0 & $\checkmark$ & 1.5288 & 0.0938 & 1.5281 & 0.0904 & 1.0192 & 1.0193 & 0.9989 & 0.9990 & 0.9999 & 1.0001 \\
        2.0 & $\checkmark$ & 1.5410 & 0.1846 & 1.5397 & 0.1772 & 1.0214 & 1.0220 & 0.9959 & 0.9963 & 0.9998 & 1.0003 \\
        3.0 & $\checkmark$ & 1.5604 & 0.2700 & 1.5582 & 0.2570 & 1.0250 & 1.0262 & 0.9912 & 0.9922 & 0.9996 & 1.0007 \\
        4.0 & $\cross$ & 1.5863 & 0.3483 & --- & --- & 1.0298 & 1.0319 & --- & --- & --- & --- \\
        \hline\hline
    \end{tabular}
    \caption{Results for rigidly rotating spheres of fixed central density $\rho_0 = 3.1$ with varying rotation parameter $v_0$. The labels of the columns are the same as the ones in Table~\ref{Tab:SpheresDensities} and the quantities with the tilde were introduced also in Sec.~\ref{Subsec:AxialSym}.}
    \label{Tab:SpheresRigidRotation}
\end{table}
\begin{figure}[htbp]
    \centering
    \includegraphics[width=0.55\textwidth]{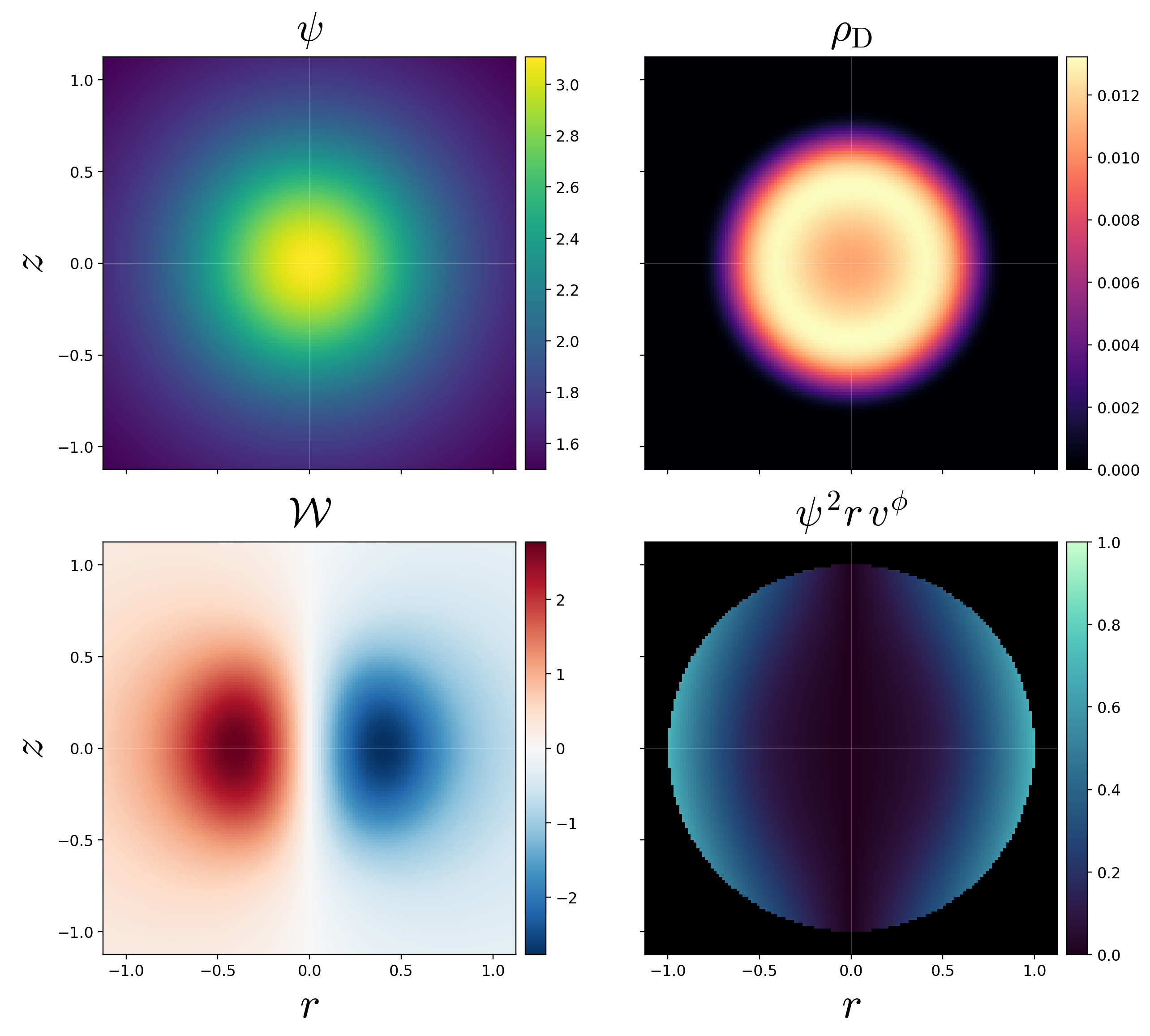}
    \caption{We represent here the solution for the rigidly rotating sphere reported in Table~\ref{Tab:SpheresRigidRotation} with $v_0 = 4$ and $\rho_0 = 3.1$. The upper left panel displays the conformal factor $\psi$, the function $\mathcal{W}$ is displayed in the lower left panel, the physical density $\rho$ is represented in the upper right panel and the physical velocity $v^{\phi} \psi^2 r $ in the lower right panel.}
    \label{Fig:RigidRotSphere}
\end{figure}

The results for the second case are reported in Table~\ref{Tab:OblateRigidRotation}, where we run simulations for oblate ellipsoids of fixed eccentricity $e = 0.99$ and density $\rho_0 = 5.1$ while varying the $v_0$ parameter in the interval $(0.0-2.5)$ in steps of $0.5$. The horizon disappears for $v_0 \geq 2.0$, again showing that, at fixed $\rho_0$, increasing the rotation removes the horizon.

\begin{table}[htbp] 
    \centering
    \setlength{\tabcolsep}{3.5pt}
    \renewcommand{\arraystretch}{1.2}
    \begin{tabular}{cccccccccccc}
        \hline\hline
        $v_0$  & AH? & $M_{\text{ADM}}$ & $\chi_{\text{ADM}}$ & $M_{\text{T}}$ & $\chi_{\mathcal{H}}$ & $\mathcal{R}_{\text{P}}^{\text{min}}$ & $\mathcal{R}_{\text{E}}^{\text{min}}$ &  $\mathcal{R}_{\text{P}}^{\text{AH}}$ & $\mathcal{R}_{\text{E}}^{\text{AH}}$ &   $\widetilde{\mathcal{R}}_{\text{P}}^{\text{AH}}$ & $\widetilde{\mathcal{R}}_{\text{E}}^{\text{AH}}$  \\
        \hline \hline
        0.0 & $\checkmark$ & 2.5082 & 0.0000 & 2.5034 & 0.0000 & 1.4405 & 1.0805 & 0.9587 & 1.1354 & 0.9587 & 1.1354 \\
        0.5 & $\checkmark$ & 2.5164 & 0.1050 & 2.5111 & 0.0959 & 1.4422 & 1.0823 & 0.9573 & 1.1353 & 0.9584 & 1.1366 \\
        1.0 & $\checkmark$ & 2.5406 & 0.2061 & 2.5332 & 0.1865 & 1.4474 & 1.0876 & 0.9533 & 1.1356 & 0.9575 & 1.1406 \\
        1.5 & $\checkmark$ & 2.5791 & 0.3000 & 2.5672 & 0.2655 & 1.4556 & 1.0960 & 0.9469 & 1.1390 & 0.9555 & 1.1494 \\
        2.0 & $\cross$ & 2.6297 & 0.3848 & --- & --- & 1.4664 & 1.1071 & --- & --- & --- & --- \\
        2.5 & $\cross$ & 2.6901 & 0.4596 & --- & --- & 1.4792 & 1.1203 & --- & --- & --- & --- \\
        \hline\hline
    \end{tabular}
    \caption{Results for rigidly rotating oblate ellipsoids of fixed eccentricity $e = 0.99$ and central density $\rho_0 = 5.1$, with varying rotation parameter $v_0$. The labels of the columns are the same as the ones in Table~\ref{Tab:SpheresRigidRotation}.}
    \label{Tab:OblateRigidRotation}
\end{table}

The results for the third case are reported in Table~\ref{Tab:ProlateRigidRotation}, where we run simulations for prolate ellipsoids of fixed eccentricity $e = 0.99$ and density $\rho_0 = 7.0$ varying the $v_0$ parameter in the interval $(0-12)$ in steps of $4$. Notably here the horizon does not disappear as we increase the velocity. We note that for $v_0=16$ the maximum physical speed would be $v_{max}=1.05 > 1$, which is superluminal and hence we do not go to higher values of $v_0$. 

\begin{table}[htbp] 
    \centering
    \setlength{\tabcolsep}{3.5pt}
    \renewcommand{\arraystretch}{1.2}
    \begin{tabular}{cccccccccccc}
        \hline\hline
        $v_0$  & AH? & $M_{\text{ADM}}$ & $\chi_{\text{ADM}}$ & $M_{\text{T}}$ & $\chi_{\mathcal{H}}$ & $\mathcal{R}_{\text{P}}^{\text{min}}$ & $\mathcal{R}_{\text{E}}^{\text{min}}$ &  $\mathcal{R}_{\text{P}}^{\text{AH}}$ & $\mathcal{R}_{\text{E}}^{\text{AH}}$ &   $\widetilde{\mathcal{R}}_{\text{P}}^{\text{AH}}$ & $\widetilde{\mathcal{R}}_{\text{E}}^{\text{AH}}$  \\
        \hline \hline
        0.0 & $\checkmark$ & 3.4439 & 0.0000 & 3.4157 & 0.0000 & 1.8415 & 0.6976 & 1.1313 & 0.8514 & 1.1313 & 0.8514 \\
        4.0 & $\checkmark$ & 3.4476 & 0.0451 & 3.4193 & 0.0434 & 1.8425 & 0.6979 & 1.1311 & 0.8511 & 1.1314 & 0.8513 \\
        8.0 & $\checkmark$ & 3.4586 & 0.0897 & 3.4298 & 0.0861 & 1.8456 & 0.6991 & 1.1306 & 0.8503 & 1.1317 & 0.8511 \\
        12.0 & $\checkmark$ & 3.4766 & 0.1331 & 3.4471 & 0.1277 & 1.8505 & 0.7009 & 1.1299 & 0.8487 & 1.1323 & 0.8504 \\
        \hline\hline
    \end{tabular}
    \caption{Results for rigidly rotating prolate ellipsoids of fixed eccentricity $e = 0.99$ and central density $\rho_0 = 7$, with varying rotation parameter $v_0$. The labels of the columns are the same as the ones in Table~\ref{Tab:SpheresRigidRotation}. }
    \label{Tab:ProlateRigidRotation}
\end{table}

Two observations regarding our results are in order now. The first thing that we notice is that, even though rotation adds energy into the configuration, for instance the ADM mass increases whenever the object starts to rotate, at fixed rescaled density it tends to make the horizons disappear. In fact, we find that for a nonrotating configuration that originally exhibits a horizon, the horizon disappears past a given point when we increase the velocity parameter. 

Second, we find that for the configurations that exhibit a horizon, we have two definitions of mass, as discussed above. One of them, the irreducible mass, is directly given in terms of the area of the horizon and the total mass receives a contribution from the rotation. For the cases reported in this section, the spin is not very high and hence we do not see a big difference between both of them. In the next subsection, we explore more rapidly spinning configurations though, where it seems to make a big difference. 

Finally, we remark that, as one would naively expect, the rotation along the $z$-axis tends to spread the horizon  in the orthogonal directions. For the prolate configurations, this means that they tend to become more spherical, explaining why the horizon seems more resilient as we increase the rotation. For the oblate (or spherical) configurations though, the configuration seems to become more oblate (it makes it deviate more from spherical symmetry) and hence the horizon quickly disappears, as one would heuristically expect.

\subsection{Differentially rotating solutions}
\label{Subsec:DifferentialRotation}
In this section we consider the same profile for the density but we consider a differentially rotating profile for $v_{\text{N}}^{\phi}$, namely
\begin{align}
    v^{\phi}_{\text{N}} (r,z) = v_0 \left(\frac{b}{r + \epsilon}\right) \left[ 1 + \kappa P \left(\xi ; \sigma \right) \right]
    \label{Eq:DiffRotProfile}
\end{align} 
where the function $P (\xi; \sigma)$ is specified in Eq.~\eqref{eq:bump}, $\epsilon$ is a parameter that enforces regularity of $v^{\phi}_{\text{N}}$ at the axis and we take it be smaller than our grid spacing. We take $\sigma = \sqrt{2}$ and $\kappa = 3.0$ while we vary the $v_0$ parameter.  

For rigid rotation, the maximum spin that we were able to find was relatively low. Our aim here is to find configurations whose dimensionless spin is as big as possible, in order to probe whether it is possible to approach $\chi_{\text{ADM}} = 1$. We also compare the Hawking and total mass as candidates for the mass relevant for the conjecture.

As mentioned in the previous section, once we invert Eq.~\eqref{Eq:vNdef} to obtain the real physical velocity, we find that $\psi$ has the effect of making the core spin slower than what we specify. Hence, in order to add more spin into the configuration without making it effectively superluminal, in particular towards the centre of the configuration, we have chosen this profile after some trial and error. As $v_0$ increases, so does the angular momentum of the configuration, and we have found that with this profile high values of $\chi_{\text{ADM}}$ can be reached. 

\begin{figure}[htbp]
    \centering
    \includegraphics[width=0.55\textwidth]{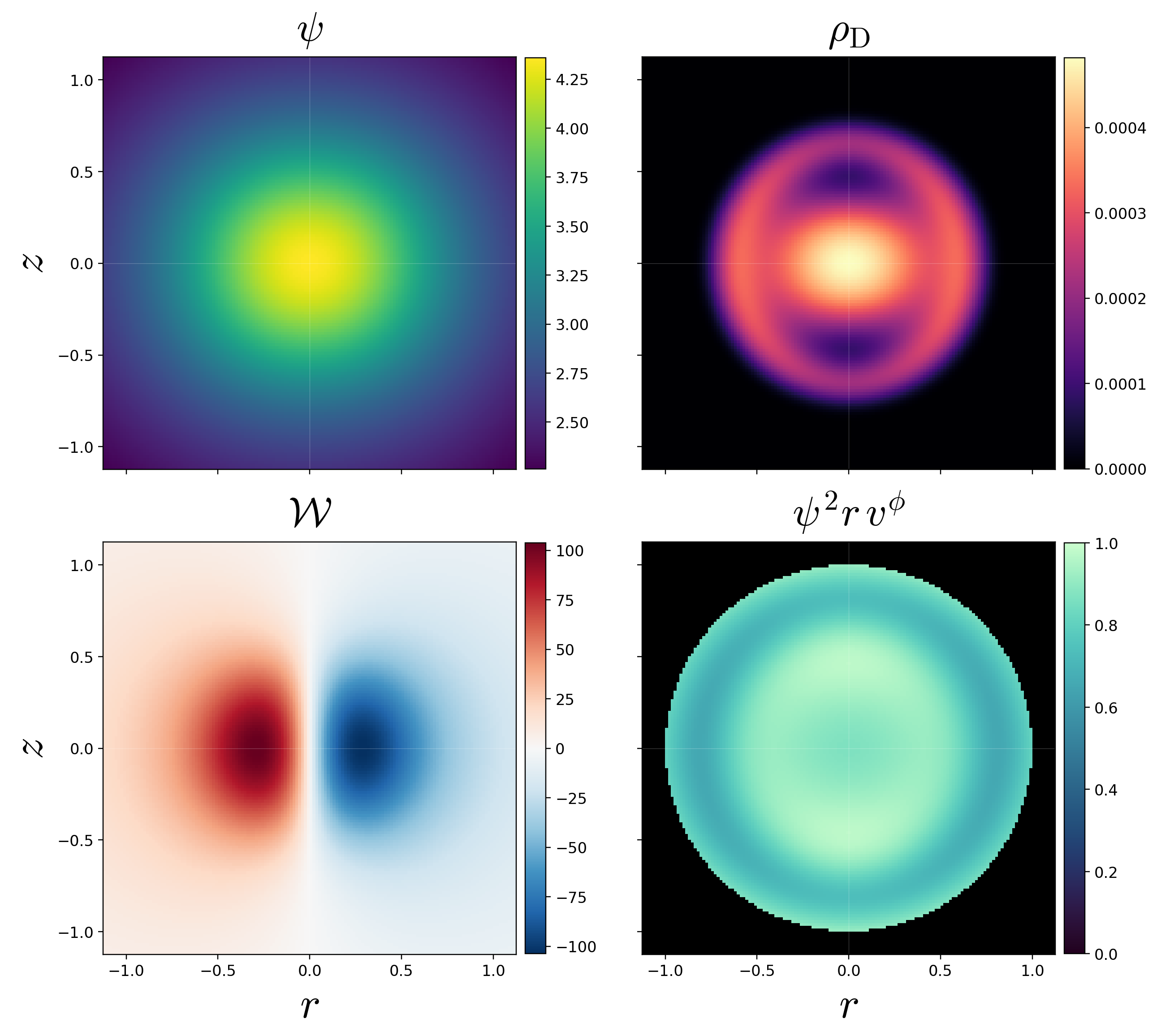}
    \caption{We represent here the solution for the differentially rotating sphere reported in Table~\ref{Tab:Sphere1} with $v_0 = 18$ and $\rho_0 = 3.1$. The upper left panel displays the conformal factor $\psi$, the function $\mathcal{W}$ is displayed in the lower left panel, the physical density $\rho$ is represented in the upper right panel and the physical velocity $v^{\phi} \psi^2 r $ in the lower right panel.}
    \label{Fig:Sphere1}
\end{figure}

We have run numerous simulations although we report here just two representative ones. We first consider a rotating sphere with $v_0 = 18.0$ and density $\rho_0 = 3.1$ for which we do not find a horizon. We plot the solution, namely $\psi$ and $\mathcal{W}$; and the physical density $\rho$ and the physical velocity $v^{\phi} \psi^2 r $ in Fig.~\ref{Fig:Sphere1}. The results are reported in Table~\ref{Tab:Sphere1}. 

\begin{table}[htbp] 
    \centering
    \setlength{\tabcolsep}{4.5pt}
    \renewcommand{\arraystretch}{1.2}
    \begin{tabular}{cccccccccccc}
        \hline\hline
        $v_0$  & AH? & $M_{\text{ADM}}$ & $\chi_{\text{ADM}}$ & $M_{\text{T}}$ & $\chi_{\mathcal{H}}$ & $\mathcal{R}_{\text{P}}^{\text{min}}$ & $\mathcal{R}_{\text{E}}^{\text{min}}$ &  $\mathcal{R}_{\text{P}}^{\text{AH}}$ & $\mathcal{R}_{\text{E}}^{\text{AH}}$ &   $\widetilde{\mathcal{R}}_{\text{P}}^{\text{AH}}$ & $\widetilde{\mathcal{R}}_{\text{E}}^{\text{AH}}$  \\
        \hline \hline
          18.0 & $\cross$ & 5.0357 & 0.9470 & --- & --- & 1.5527 & 1.6109 & --- & --- & --- & --- \\
        \hline\hline
    \end{tabular}
    \caption{Results for a differentially rotating sphere of central density $\rho_0 = 3.1$ and rotation parameter $v_0 = 18.0$. The labels of the columns are the same as the ones in Table~\ref{Tab:SpheresRigidRotation}. Note that the contained mass here is $M_{\text{cont}}  = 2.44897$, just less than half of the ADM mass. Notice that the ADM dimensionless spin is below one. }
    \label{Tab:Sphere1}
\end{table}

\begin{figure}[htbp]
    \centering
    \includegraphics[width=0.55\textwidth]{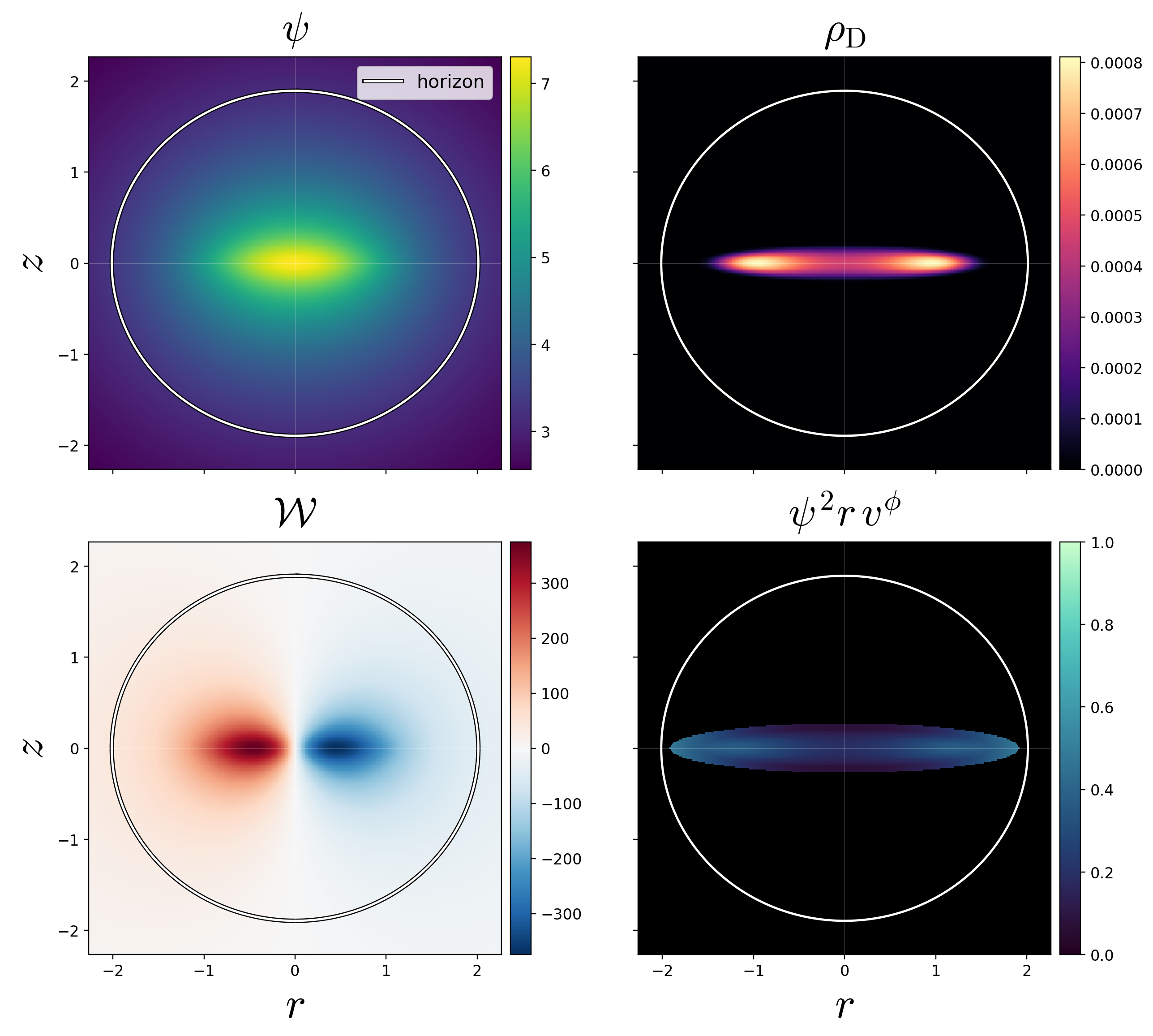}
    \caption{We represent here the solution for the differentially rotating sphere reported in Table~\ref{Tab:Pancake1} with $v_0 = 10$ and $\rho_0 = 10$ and $e = 0.99$. In the upper left panel we display the conformal factor $\psi$, the function $\mathcal{W}$ is displayed in the lower left panel, the physical density $\rho$ is represented in the upper right panel and the physical velocity $v^{\phi} \psi^2 r$ in the lower right panel. The white curve represents the location of the apparent horizon.}
    \label{Fig:Pancake1}
\end{figure}

We also report here a rotating oblate configuration for which we have considered $v_0 = 10.$, $e = 0.99$ and $\rho_0 = 10.0$ for which we do find a horizon. We plot the solution, namely $\psi$ and $\mathcal{W}$; and the physical density $\rho$ and the physical velocity $v^{\phi} \psi^2 r $ in Fig.~\ref{Fig:Pancake1}. The results are reported in Table~\ref{Tab:Pancake1}. In this example we clearly see from $\mathcal{R}^{\text{AH}}_{E}$ and $\widetilde{\mathcal{R}}^{\text{AH}}_{E}$ that we need to consider the total mass $M_T$, since otherwise we see violations of the hoop conjecture by up to $6 \%$. The inverse hoop conjecture~\cite{Hod2020} is also satisfied by this configuration and by all the rigidly rotating configurations with an apparent horizon, i.e., $\max\big(\widetilde{\mathcal{R}}^{\text{AH}}_{\text{P}}, \widetilde{\mathcal{R}}^{\text{AH}}_{\text{E}}\big) \geq 1$ in all cases.

\begin{table}[htbp] 
    \centering
    \setlength{\tabcolsep}{4.5pt}
    \renewcommand{\arraystretch}{1.2}
    \begin{tabular}{cccccccccccc}
        \hline\hline
        $v_0$  & AH? & $M_{\text{ADM}}$ & $\chi_{\text{ADM}}$ & $M_{\text{T}}$ & $\chi_{\mathcal{H}}$ & $\mathcal{R}_{\text{P}}^{\text{min}}$ & $\mathcal{R}_{\text{E}}^{\text{min}}$ &  $\mathcal{R}_{\text{P}}^{\text{AH}}$ & $\mathcal{R}_{\text{E}}^{\text{AH}}$ &   $\widetilde{\mathcal{R}}_{\text{P}}^{\text{AH}}$ & $\widetilde{\mathcal{R}}_{\text{E}}^{\text{AH}}$  \\
        \hline \hline
        10.0 & $\checkmark$ & 13.122 & 0.922 & 12.938 & 0.949 & 2.879 & 2.052 & 0.790 & 0.858 & 0.973 & 1.058 \\
        \hline\hline
    \end{tabular}
    \caption{Results for a differentially rotating oblate spheroid with parameters $\rho_0 = 10.0$, $v_0 = 10.0$ and $e = 0.99$. The labels of the columns are the same as the ones in Table~\ref{Tab:SpheresRigidRotation}. In this example the hoop conjecture is satisfied with the total mass but violated with the irreducible mass.}
    \label{Tab:Pancake1}
\end{table}

The conclusions that we extracted from the previous section are also supported by all the examples of differentially rotating $v^{\phi}_{\text{N}}$ that we have run, even though we only report here two particularly relevant situations. Interestingly enough, among all the simulations that we have run, we find that all the examples for which we get close to $\chi \sim 1$, an effective superluminal behaviour arises. This perfectly aligns with the discussion from Sec.~\ref{Subsec:OverSpinning}, as we do not expect to be able to find configurations with $\chi \geq 1$ for physically realistic matter content, which in this case means subluminally rotating dust, even if they spin very fast. Hence, our results support such conjecture which we recall, up to now has only been proved in vacuum and some plausibility arguments have been given for fluid spheroids.

Regarding the hoop conjecture, the two mass definitions now give noticeably different results. For the configuration in Table~\ref{Tab:Pancake1}, with $\chi_{\mathcal{H}} = 0.949$, using the irreducible mass instead of $M_T$ increases the horizon ratios by a factor $M_T/M_{\mathcal{H}} \simeq 1.23$. The equatorial ratio then exceeds unity, $\widetilde{\mathcal{R}}^{\text{AH}}_{\text{E}} = 1.058$, whereas $\mathcal{R}^{\text{AH}}_{\text{E}} = 0.858$. This excess is smaller than the deviations produced by eccentricity alone in the nonrotating case (up to $\simeq 14\%$), so on its own it is not conclusive. However, the shift between the two masses depends only on the horizon spin, through Christodoulou's formula, and grows monotonically with it, reaching a factor $\sqrt{2}$ for extremal Kerr, Eq.~\eqref{Eq:KerrExtremalHoop}. This indicates that, for rotating configurations, the mass entering the hoop conjecture should include the rotational energy, in agreement with the Kerr argument above.

We expect the difference to become more pronounced at higher spins. Within our setup, however, larger values of $\chi_{\mathcal{H}}$ cannot be reached without the dust becoming superluminal. Exploring matter models that allow faster rotation would be a natural way to sharpen this conclusion.

For the \emph{only if} part of the conjecture, the role of rotation has, to our knowledge, only been discussed in Refs.~\cite{Senovilla2007,Gibbons2009,Cvetic2011} and mainly for stationary black holes; our results agree with that discussion. For the \emph{if} part, comparing the minimal circumferences enclosing the matter with the contained mass $M_{\text{cont}}$, Eq.~\eqref{Eq:ContM}, we find the conjecture satisfied in all horizonless configurations. We stress, however, that this conclusion depends on how the rotational field energy is assigned: for the configuration of Table~\ref{Tab:Sphere1}, $M_{\text{cont}} \simeq 0.49\, M_{\text{ADM}}$, and using $M_{\text{ADM}}$ instead would give $\mathcal{R}^{\text{min}}_{\text{P,E}} \simeq 0.76$ and $0.78$, which would constitute a violation of the hoop conjecture. Unlike the nonrotating case, where the ADM and the contained mass agree, here we see again that it is needed to account properly for the rotational energy contained inside the surface for the conjecture to hold.

\section{Conclusions and discussion}
\label{Sec:Conclusions}
We have constructed initial data for axisymmetric, rotating dust configurations by solving the GR constraint equations on a maximal slice, assuming a conformally flat 3-metric and no radial momentum. To avoid the nonexistence and nonuniqueness of solutions that arise when the physical density and angular velocity are prescribed directly, we specify suitably rescaled sources, and we express the ADM mass and angular momentum as integrals over these sources and the resulting fields.

In this setting the extrinsic curvature drops out of the apparent-horizon condition, so that apparent horizons are minimal surfaces of the 3-metric even for rotating configurations. We locate them with three independent algorithms, which agree well with one another, and we compute the horizon area, angular momentum, and Christodoulou mass from surface integrals.

Building on the work of Nakamura, Shapiro, and Teukolsky~\cite{Nakamura1988}, we have used these data to test the hoop conjecture and discuss its connection to weak cosmic censorship. The main novelty is the inclusion of rotation. After reviewing the difficulties in formulating the conjecture precisely, we have shown how they can be addressed in our setup through a specific choice of slicing, of the surfaces on which circumferences are measured, and of the quasilocal mass.

For the nonrotating configurations we find a good agreement with the results of Ref.~\cite{Nakamura1988}, in particular that it does not seem possible to determine a universal coefficient for the \emph{if} part of the conjecture. However, for configurations exhibiting a horizon the inequality is satisfied within $14\%$ of error in all the cases examined, which corresponds to the \emph{only if} part of the conjecture.

For rotating dust, when a horizon is present, two different masses can be associated with the apparent horizon, namely the Hawking (irreducible) mass which is roughly speaking the area of the horizon; or the total mass which is obtained using Christodoulou's formula and includes a contribution from the angular momentum of the horizon. Our conclusions are similar to the ones advanced in Ref.~\cite{Senovilla2007}, namely that the Hawking mass does not seem a good candidate for the hoop conjecture.

For the \emph{only if} part, using the irreducible mass of the horizon leads to violations of up to $\simeq 6\%$ for our most rapidly spinning configuration, and the shift relative to the Christodoulou mass grows with spin, as for Kerr. With the Christodoulou mass, the horizon ratios remain within $\simeq 14\%$ of unity, as in the nonrotating case. We therefore conclude that the quasilocal mass entering the hoop conjecture should include a rotational contribution. Configurations with higher spins, which cannot be reached within our setup without the dust becoming superluminal, would make this difference more pronounced. On the other hand, Hod's inverse hoop conjecture~\cite{Hod2020} remarkably holds in all the cases that we have studied. Hence, it would be interesting to investigate more exotic configurations that may push the conjecture to its limits.

A similar conclusion holds for the \emph{if} part. Using the contained mass $M_{\text{cont}}$, i.e., the energy enclosed by the matter surface including its rotational contribution, the conjecture is satisfied in all our horizonless configurations. Using the ADM mass, which also includes the rotational field energy outside the matter and coincides with $M_{\text{cont}}$ only without rotation, the minimal hoops can be up to $\simeq 25\%$ smaller than $4\pi M_{\text{ADM}}$ (Table~\ref{Tab:Sphere1}). Both tests are thus consistent with a single prescription: the mass in the hoop inequality is the energy enclosed by the surface on which the hoop is measured, including its rotational part.

In this article, we have not considered non-axially symmetric configurations, although it is a very interesting line of future work, considering our findings here. In Ref.~\cite{Senovilla2007}, it is argued that the right-hand side of the hoop inequality should include an integral of a geometric quantity over the horizon, which is closely related to angular momentum and gravitational radiation~\cite{Epp2000}. This is conceptually similar to the conclusion reached here, since, in the axisymmetric case, this integral is directly related to the angular momentum defined geometrically through the axial Killing field. We consider it worthwhile to investigate this point further and scrutinize the role of this quantity as a possible measure of the rotational energy of the horizon that should be accounted for in the hoop conjecture.

Another interesting follow-up would be to consider initial data which contain some gravitational waves, i.e., the transverse traceless part of the extrinsic curvature that is undetermined by constraints. It would be informative to check whether a similar definition of the mass contained in a region can be given, and whether it satisfies the conjecture or not.

We also have tried to find configurations in which the quantity $\chi_{\text{ADM}}$ approaches one, but we have not been able to identify any physical configuration for which we can reach such value. In general, we have found out that every time we specified a profile that was potentially able to reach such value, the solution becomes unphysical, i.e., dust seems to be moving superluminally and violating energy conditions. This supports the expectation that $\chi_{\text{ADM}} \leq 1$ for configurations close to horizon formation (see Sec.~\ref{Subsec:OverSpinning}). A configuration violating this bound and collapsing to a black hole would lead to a superextremal end state, in tension with weak cosmic censorship.

Our results also have potential implications for horizonless exotic compact objects~\cite{Cardoso:2019rvt}. At fixed rescaled density, increasing the rotation can remove the apparent horizon even though the ADM mass grows (Tables~\ref{Tab:SpheresRigidRotation} and~\ref{Tab:OblateRigidRotation}). Moreover, the differentially rotating configuration of Table~\ref{Tab:Sphere1} has no apparent horizon despite an ADM mass more than three times larger than that of the nonrotating sphere at the threshold of horizon formation (Table~\ref{Tab:SpheresDensities}). This suggests that rotation, and differential rotation in particular, may help very compact configurations avoid horizon formation. Since rotation in exotic compact objects beyond GR is poorly understood, this deserves further investigation, for instance by comparing configurations with the same ADM mass and angular momentum but different rotation profiles.

Finally, the time evolution of the configurations that we have constructed is also an interesting question, in particular regarding their interplay with weak cosmic censorship. For the nonrotating ones, it was first argued that the evolution of some horizonless configurations led to unbounded curvature invariants without the formation of a horizon~\cite{Shapiro1991,Yoo2016}, seeming to suggest that a naked singularity could form. However, more recent analysis~\cite{East2019,Ames2023}, suggest that an apparent horizon does form (we recall that apparent horizons are slicing-dependent~\cite{Wald1991}), and all the matter content is swallowed by a black hole, with a strong emission of gravitational waves. It would be interesting to understand whether this picture changes for initially rotating configurations, especially since, at fixed rescaled density, rotation tends to remove the apparent horizon in our initial data. It would also be interesting to compare the gravitational-wave emission from the collapse of rotating configurations with those found in the nonrotating case.

%

%--------------------------------------------------------------------
\section*{Acknowledgements}
\addcontentsline{toc}{section}{Acknowledgements}
%--------------------------------------------------------------------

We would like to thank Alexandru Dima, Andrea Sanna, Carlos Barcel\'o, and Alejandro Jim\'enez Cano for helpful discussions. We also want to thank Miguel S\'anchez and Francisco Torralbo for useful correspondence. We are grateful to Shahar Hod for bringing the inverse hoop conjecture to our attention. The authors are supported by the MUR FIS2 Advanced Grant ET-NOW (CUP:~B53C25001080001) and by the INFN TEONGRAV initiative. GGM acknowledges funding from the project Grant No. PID2023-149018NB-C43 funded by MCIN/AEI/10.13039/501100011033. 

%--------------------------------------------------------------------
\newpage
%--------------------------------------------------------------------

%--------------------------------------------------------------------
\appendix
%--------------------------------------------------------------------

%--------------------------------------------------------------------
\section{Lichnerowicz decomposition}
\label{App:ExtrinsicCurvature}
%--------------------------------------------------------------------

The only nonvanishing Christoffel symbols are $\Gamma^{r}_{\ \phi \phi} = - r$ and $\Gamma^{\phi}_{\ r \phi} = \Gamma^{\phi}_{\ \phi r } = \frac{1}{r}$. The expression for the Lichnerowicz acting on $W^{i}$ is
\begin{align}
    \left( \boldsymbol{\tilde{L}} W \right)^{ij} := \tilde{\gamma}^{ik} \tilde{D}_k W^j +\tilde{\gamma}^{jk} \tilde{D}_k W^i - \frac{2}{3} \tilde{\gamma}^{ij} \tilde{D}_k W^k.
\end{align}
The divergence vanishes for $W^k = W^{\phi} \delta^{k}_{\ \phi}$
\begin{align}
    \tilde{D}_k W^k = \partial_k W^k + \Gamma^{k}_{\ k j} W^j = \partial_{\phi} W^{\phi} + \Gamma^{k}_{\ k \phi} W^\phi = 0.
\end{align}
Computing the covariant derivatives:
\begin{align}
     \tilde{\gamma}^{ik} \tilde{D}_k W^j = \tilde{\gamma}^{ik} \partial_k W^j + \tilde{\gamma}^{ik} \Gamma^{j}_{\ k l} W^l,
\end{align}
the only nonvanishing ones are:
\begin{align}
    & \tilde{D}_r W^{\phi} = \partial_r W^{\phi} + \frac{1}{r} W^{\phi}, \\
    & \tilde{D}_z W^{\phi} = \partial_z W^{\phi}, \\
    & \tilde{D}_{\phi} W^{r} = - r W^{\phi}.
\end{align}
This means that the only nonvanishing components of $\left( \boldsymbol{\tilde{L}} W \right)^{ij}$ are:
\begin{align}
    & \left( \boldsymbol{\tilde{L}} W \right)^{r \phi} =\left( \boldsymbol{\tilde{L}} W \right)^{\phi r} = \partial_r W^{\phi}, \\
    & \left( \boldsymbol{\tilde{L}} W \right)^{z \phi} = \left( \boldsymbol{\tilde{L}} W \right)^{ \phi z } = \partial_z W^{\phi}. 
\end{align}
In the equations it appears $   \tilde{D}_j \left( \boldsymbol{\tilde{L}} W \right)^{ij} $, so we compute:
\begin{align}
    \tilde{D}_k \left( \boldsymbol{\tilde{L}} W \right)^{ij} = \partial_k \left( \boldsymbol{\tilde{L}} W \right)^{ij} + \Gamma^{i}_{\ kl} \left( \boldsymbol{\tilde{L}} W \right)^{lj} + \Gamma^{j}_{\ kl} \left( \boldsymbol{\tilde{L}} W \right)^{il}, 
\end{align}
so contracting $k = j$:
\begin{align}
    \tilde{D}_j \left( \boldsymbol{\tilde{L}} W \right)^{ij} = \partial_j \left( \boldsymbol{\tilde{L}} W \right)^{ij} + \Gamma^{i}_{\ jl} \left( \boldsymbol{\tilde{L}} W \right)^{lj} + \Gamma^{j}_{\ jl} \left( \boldsymbol{\tilde{L}} W \right)^{il}.
\end{align}
Computing the components $r$ and $z$ of the equation, they identically vanish. We get for the $\phi$ component: 
\begin{align}
    & \tilde{D}_j \left( \boldsymbol{\tilde{L}} W \right)^{\phi j} = \partial_j \left( \boldsymbol{\tilde{L}} W \right)^{\phi j} + \Gamma^{\phi}_{\ jl} \left( \boldsymbol{\tilde{L}} W \right)^{lj} + \Gamma^{j}_{\ jl} \left( \boldsymbol{\tilde{L}} W \right)^{\phi l} = \nonumber \\
    & = \partial_r  \left( \boldsymbol{\tilde{L}} W \right)^{\phi r}  + \partial_z  \left( \boldsymbol{\tilde{L}} W \right)^{\phi z} +\Gamma^{\phi}_{\ r \phi}  \left( \boldsymbol{\tilde{L}} W \right)^{r \phi} + \Gamma^{\phi}_{\ \phi r}  \left( \boldsymbol{\tilde{L}} W \right)^{\phi r} + \Gamma^{\phi}_{\ \phi r}  \left( \boldsymbol{\tilde{L}} W \right)^{ \phi r } = \nonumber \\
    & \left(\partial_r^2 + \partial_z^2 + \frac{3}{r} \partial_r \right) W^{\phi} .  
\end{align}
The term $\left( \boldsymbol{\tilde{L}} W \right)^{ij} \left( \boldsymbol{\tilde{L}} W \right)_{ij}$:
\begin{align}
    & \tilde{\gamma}_{i l} \tilde{\gamma}_{jk} \left( \boldsymbol{\tilde{L}} W \right)^{ij} \left( \boldsymbol{\tilde{L}} W \right)^{l k} = 2 \tilde{\gamma}_{rr} \tilde{\gamma}_{\phi \phi} \left[ \left( \boldsymbol{\tilde{L}} W \right)^{r \phi} \right]^2 + 2 \tilde{\gamma}_{zz} \tilde{\gamma}_{\phi \phi} \left[ \left( \boldsymbol{\tilde{L}} W \right)^{z \phi} \right]^2 \nonumber  \\
    & = 2r^2 \left[\left( \partial_r W^{\phi} \right)^2 + \left( \partial_z W^{\phi} \right)^2 \right] . 
\end{align}
%
%--------------------------------------------------------------------
\section{Boundary conditions for the shooting method}
\label{App:HorizonAppendix}
Consider the axisymmetric horizon surface $R(\theta)$ that we are targeting to compute with the shooting method. It should satisfy Eq.~\eqref{Eq:Horizon}, hence if we specify the value of the function $R(\theta_0)$ and its derivative $R'(\theta_0)$ at a point, we can compute all of its higher derivatives. In particular, we know that regularity of the surface at the pole $\theta = 0$ enforces $R'( 0 ) = 0$. In fact, the function should be even and hence all the odd order derivatives should identically vanish at the pole. This automatically provides the following Taylor expansions near the pole 
\begin{align}
    & R(\theta) = R_{\text{p}} + \frac{1}{2} R^{\prime \prime}_{\text{p}} \theta^2 + \order{\theta^4} \label{Eq:PoleRShoot} \\
    & R' (\theta) = R^{\prime \prime}_{\text{p}} \theta + \order{\theta^3}, \label{Eq:PoleRprimeShoot}
\end{align}
where $R_{\text{p}}$ is the (unknown at this point) value of the radius of the horizon at the pole and the second derivative is completely determined from Eq.~\eqref{Eq:Horizon}:
\begin{align}
    R^{\prime \prime}_{\text{p}} = R_{\text{p}} + 2 R_{\text{p}}^2 \partial_R \log \psi \left( R_{\text{p}}, 0 \right).
\end{align}
\emph{Mutatis mutandis}, reflection symmetry allows us to get a similar expansion around the equator where $R_{\text{eq}}$ is now the
\begin{align}
    & R(\theta) = R_{\text{eq}} + \frac{1}{2} R^{\prime \prime}_{\text{eq}} ( \theta - \pi/2)^2 + \order{(\theta - \pi/2)^4} \label{Eq:EquatorRShoot}\\
    & R' (\theta) = R^{\prime \prime}_{\text{eq}} (\theta - \pi/2) + \order{(\theta - \pi/2)^3}, \label{Eq:EquatorRprimeShoot}
\end{align}
with
\begin{align}
    R^{\prime \prime}_{\text{eq}} = 2R_{\text{eq}} + 4 R_{\text{eq}}^2 \partial_R \log \psi \left( R_{\text{eq}}, \frac{\pi}{2} \right).
\end{align}
\section{Near horizon approximation} 
\label{App:NearHorizon}
We consider here small deformations of a surface that is close to the Schwarzschild horizon and compute its expansion perturbatively, $\Theta^{+}$, to illustrate how the relaxation algorithms converge to the horizon surface. Consider a constant time slice of the Schwarzschild geometry expressed in isotropic coordinates
\begin{align}
    \dd s^2 = -  \left( \frac{1 - M/(2R)}{1 + M/(2R)} \right)^2 \dd t^2 + \left( 1 + \frac{M}{2 R} \right)^4 \left( \dd R^2 + R^2 \dd \theta^2 + R^2 \sin^2 \theta \dd \phi^2 \right)
\end{align}
and define an embedded 2D surface $h(\theta,\phi)$, which is quite close to the Schwarzschild horizon, as
\begin{align}
    \mathcal{F} = R - \frac{M}{2} - h(\theta,\phi) = 0,
\end{align}
where $h$ is small. In these coordinates, the horizon sits at $R = M/2$. The unit normal to $h$ is
\begin{align}
    \hat{{s}}_i \;&=\; 
    \frac{1}{\sqrt{\gamma^{ij}\partial_i \mathcal{F} \partial_j \mathcal{F}}}
    \begin{pmatrix}
        1 \\ -\frac{\partial h}{\partial \theta} \\ -\frac{\partial h}{\partial \phi}
    \end{pmatrix},
\end{align}
The isotropic metric is diagonal, $\gamma_{ij}=\psi^4{\rm diag}(1,R^2,R^2\sin^2\theta)$, with $\psi = 1 + M/(2R)$. The off-diagonal terms vanish identically and we obtain
\begin{align}
    |\nabla\mathcal{F}| \;=\;  \psi^{-2} + \mathcal{O}(h^2) ,
\end{align}
so the angular gradients of $h$ drop out of the norm altogether. We can write
\begin{align}
     R = \frac{M}{2}\left(1 + \frac{2}{M}h\right) , 
\end{align}
or equivalently 
\begin{align}
     \frac{2}{M} R = 1 + \frac{2}{M} h.
\end{align}
The conformal factor transforms and its derivatives can also be expanded as
\begin{align}
    \psi &= 1 + \frac{M}{2R} = 2\left(1 - \frac{h}{M} \right) + \mathcal{O}(h^2), \\
    \partial_R \psi &= -\frac{M}{2R^2} = - \frac{2}{M}+ \frac{8h}{M^2} + \mathcal{O}(h^2).
\end{align}
The null expansion derived earlier is, 
\begin{align}
    \Theta^+ = -\frac{1}{|\nabla \mathcal{F}|} \bigg[ \nabla^2 h(\theta,\phi) - \nabla^2 R + \frac{ (\partial_i \mathcal{F}) \partial^i|\nabla\mathcal{F}|}{|\nabla \mathcal{F}|}  \bigg]. 
\end{align}
Evaluating each term gives
\begin{align}
    \nabla^2 h &= \frac{\psi^{-4}}{R^2}\,\mathring{\Delta} h
        = \frac{1}{4M^2}\,\mathring{\Delta} h , \\
    \nabla^2 R &= \psi^{-4}\left(\frac{2}{R} + \frac{2\partial_R\psi}{\psi}\right)
        = \frac{1}{8M} + \frac{3h}{8M^2} , \\
    \frac{(\partial_i \mathcal{F})\, \partial^i |\nabla\mathcal{F}|}
         {|\nabla\mathcal{F}|}
        &= -\frac{2\partial_R \psi}{\psi^{5}}
        = \frac{1}{8M} + \frac{h}{8M^2} ,
\end{align}
with $\mathring{\Delta} = \partial_\theta^2 + \cot\theta\,\partial_\theta + \sin^{-2}\theta\,\partial_\phi^2$ the unit-sphere Laplacian. The $\mathcal{O}(1)$ pieces cancel between the last two equations, as they must for $h=0$ to be the horizon, leaving
\begin{align}
    \Theta^+ = \frac{1}{M^2}\Big( h - \mathring{\Delta} h \Big) + \mathcal{O}(h^2),
\end{align}
which is a remarkably simple expression.

Finally, we can decompose $\Theta^{+}$ as
\begin{align}
    \Theta^{+} (\theta, \phi) = \sum_{\ell = 0}^{\infty} \sum_{m = - \ell}^{\ell} \Theta^{+}_{\ell m} Y_{\ell m} (\theta, \phi),
\end{align}
where $Y_{\ell m} (\theta, \phi)$ are the spherical harmonics, to find the following to leading order in $h_{\ell m}$ 
\begin{align}
    \Theta^+ (\theta, \phi) = \sum_{\ell = 0}^{\infty} \lambda_\ell h_{\ell m} \sum_{m = - \ell}^{\ell} \, Y_{\ell m} (\theta,\phi), \qquad \lambda_\ell = \frac{\ell^2 + \ell + 1}{M^2} .
\end{align}
Every eigenvalue is strictly positive, so applying the parabolic relaxation with this surface as an initial seed, we see that all the angular perturbations damp and we conclude that the horizon is a genuine attractor. The spectrum is bounded below by $\lambda_0 = 1/M^2$, so convergence never becomes arbitrarily slow, but grows as $\ell^2$, which is actually the stiffness that sets the maximum time step allowed.

\section{Convergence and grid setup}
\label{App:Convergence}

All the configurations in this paper are solved on the domain $0 \le [r, z] \le L$ with $L = 64$, using ten grids from $8^2$ cells ($\Delta_0 = 8$) to $4096^2$ cells ($\Delta_9 = 1/64$). The smallest source we consider is the $e = 0.99$ oblate spheroid; $\Delta_9 = 1/64$ puts 17 cells across the semi-minor axis of the matter, or 34 across its diameter. Each solve takes about 20 V-cycles to reach $\mathcal{E} < 10^{-9}$, around a minute or less on a four-core desktop.

In order to test the validity of the constraint solver, we tested it against the following cases,
\begin{itemize}
    \item The first of these is for a nonrotating ($v^{\phi} = \mathcal{W}=0$) constant density ($\rho_{\text{N}}$) sphere of radius $a$. The total mass is $M=4\pi a^3\rho_{\text{N}} /3$. In this case, Eq.\eqref{Eq:PsiRescequation} reduces to a linear Poisson equation of flat space,
\begin{align} \label{eq:poisson_simple}
    \nabla^2 \psi = -2 \pi \rho_{\text{N}}
\end{align}
which can be solved analytically, 
\begin{equation} 
    \psi =
        \begin{cases}
            1 + \dfrac{M}{2R}, & R > a, \\[6pt]
            1 + \frac{M}{4a} \left(3 -  \frac{R^2}{a^2} \right), & R \leq a.
        \end{cases}
        \label{Eq:ConstantDensityTest}
\end{equation}
\item Secondly, working directly with the physical source $\rho_{\text{D}} = \rho_{\text{N}} \psi^{-5}$ transforms  
Eq.~\eqref{eq:poisson_simple} to
\begin{align}
    \nabla^2 \psi = - 2 \pi \psi^5 \rho_{\text{D}} ;
\end{align}
for sub-critical choices of $\rho_{\text{D}}$, solutions for $\psi$ in spherical symmetry are given in ~\cite{Baumgarte2006}.
\item Thirdly, removing spherical symmetry for ellipsoidal matter configurations, but returning to the Newtonian source definition $\rho_{\text{N}}$, Nakamura et al. \cite{Nakamura1988} give analytic solutions for $\psi(r,z)$. 
\end{itemize}

\begin{figure}[htbp]
    \centering
    \includegraphics[width=0.8\textwidth]{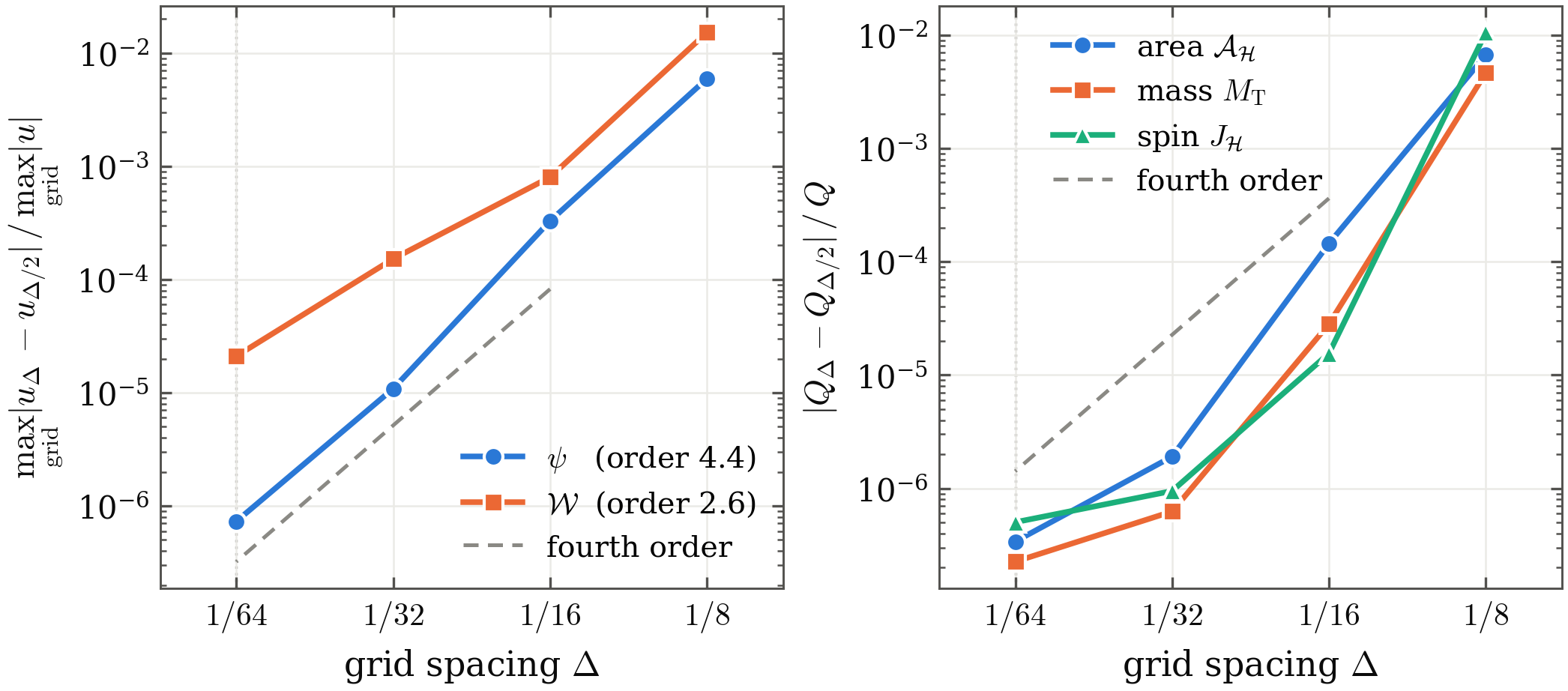}
    \caption{Convergence test for the differentially rotating oblate configuration of Table~\ref{Tab:Pancake1} for grids from $512^2$ to $8192^2$ cells. (Left) Maximum changes of $\psi$ and $\mathcal{W}$ between successive resolutions, normalized to to the maximum value of each field. (Right) Relative change of the horizon area, Christodoulou mass and angular momentum. For the grid used in this paper ($\Delta$ = 1/64) the horizon quantities change by less than $10^{-6}$, and $\psi$ and $\mathcal{W}$ by less than $10^{-6}$ and $3\times10^{-5}$, respectively.}
    \label{fig:convergence}
\end{figure}

In addition to these comparisons to known solutions, internal self consistency checks have been carried out. For instance, we have tried different values for the outer boundary distance $L$ and verified that the numerical solution does not change by a relevant amount. Furthermore, we have compared different types of boundary conditions for $\psi$ in the nonrotating case, namely the Robin-type boundary conditions of Eq.~\eqref{eq:bdry_dirichlet_psi} and the Dirichlet type which explicitly sets \mbox{$\psi = 1 + M/(2R)$} where $M$ represents the integral of $\rho_{\text{N}}$. In all the cases, we have found that the solutions converge well, as shown in Fig.~\ref{fig:convergence}.

\newpage

%--------------------------------------------------------------------

%------------------------------------------------
\bibliographystyle{JHEP}
\bibliography{references}
%------------------------------------------------

%--------------------------------------------------------------------
\end{document}